\documentclass[pra,superscriptaddress,twocolumn,floatfig,showpacs,floatfix]{revtex4-1}

\usepackage{graphicx,color}
\usepackage{amsmath,amssymb,bm}
\usepackage{braket}	

\usepackage{mathtools}
\usepackage{tikz}
\usetikzlibrary{calc,arrows.meta,positioning}
\usetikzlibrary{arrows,positioning}
\usetikzlibrary{shapes,snakes}
\usepackage{float}
\usepackage[plainpages=false,pdfpagelabels,colorlinks=true,linkcolor=blue,urlcolor=blue,citecolor=blue,pdftitle={Finite-temperature optical conductivity with density-matrix renormalization group methods for the Holstein polaron and bipolaron with dispersive phonons},pdfauthor={},pdfdisplaydoctitle=true,pdfduplex=DuplexFlipLongEdge]{hyperref}

\definecolor{darkred}{rgb}{0.90,0,0}
\definecolor{darkgreen}{rgb}{0,0.60,.2}
\definecolor{darkblue}{rgb}{0,0,1}
\definecolor{grey}{cmyk}{0,0,0,0.25}
\definecolor{orange}{cmyk}{0,0.6,1,0}

\usepackage{physics}
\usepackage{mathrsfs}
\usepackage{bbold}
\begin{document}
\title{
Finite-temperature optical conductivity with density-matrix renormalization group methods for the Holstein polaron and bipolaron with dispersive phonons
}

\author{David Jansen}
\affiliation{Institut f\"ur Theoretische Physik, Georg-August-Universit\"at G\"ottingen, D-37077 G\"ottingen, Germany}

\author{Janez Bon\v{c}a}
\affiliation{J. Stefan Institute, 1000 Ljubljana, Slovenia}
\affiliation{Faculty of Mathematics and Physics, University of Ljubljana, 1000 Ljubljana, Slovenia}

\author{Fabian Heidrich-Meisner}
\affiliation{Institut f\"ur Theoretische Physik, Georg-August-Universit\"at G\"ottingen, D-37077 G\"ottingen, Germany}

\begin{abstract}
A comprehensive picture of polaron and bipolaron physics is essential to understand the optical absorption spectrum in many materials with electron-phonon interactions. In particular, the finite-temperature properties are of interest since they play an important role in many experiments. Here, we combine the parallel two-site time-dependent variational principle algorithm (p2TDVP) with local basis optimization (LBO) and purification to calculate time-dependent current-current correlation functions. From this information, we extract the optical conductivity for the Holstein polaron and bipolaron with dispersive phonons at finite temperatures. For the polaron in the weak and intermediate electron-phonon coupling regimes, we analyze the influence of phonon dispersion relations on the spectra.  For strong electron-phonon coupling, the known result of an asymmetric Gaussian is reproduced for a flat phonon band. For a finite phonon bandwidth, the center of the Gaussian is either shifted to larger or smaller frequencies, depending on the sign of the phonon hopping. We illustrate that this can be well understood by considering the Born-Oppenheimer surfaces.  Similar behavior is seen for the bipolaron for strong coupling. For the bipolaron with weak and intermediate coupling strengths and a flat phonon band, we obtain two very different spectra. The latter also has a temperature-dependent resonance at a frequency below the phonon frequency.
\end{abstract}

\maketitle
\section{Introduction}
\label{sec:intro}
The interaction of an electron with lattice vibrations leads to the formation of a quasi-particle called a polaron, see, e.g., Ref.~\cite{franchini_21} for a review. Such a quasi-particle formation is of both theoretical and experimental interest since it is responsible for many material properties, e.g., in manganites~\cite{millis_95,Lanzara_98,Jooss_07,Schramm_2008}, lead halide, and hybrid perovskites~\cite{miyata_17,cortecchia_17,cinquanta_19,Ghosh_2020}. Since one way to experimentally study polaron physics is optical absorption spectroscopy, key insights into such materials can result from a better understanding of the optical excitation spectrum of these quasi-particles. These optical excitations allow for, e.g., the detection of small polarons, see, e.g., Refs.~\cite{yoon_98,Quijad_98,mildner_15}. In several cases, theory and experiments were successfully combined to extract information about polarons in manganites, see, e.g., Refs.~\cite{kim_96,yoon_98,hartinger_06,mildner_15}.

Experiments are inevitably conducted at finite temperatures. This calls for a comprehensive understanding of the temperature dependence of the polaron optical conductivity.  For example, in Ref.~\cite{machiada_98}, absorption spectra were used to investigate films of doped manganites by decreasing temperature, with a transition from a paramagnetic into a ferromagnetic phase. This led to the detection of a transition from small to large polarons. Furthermore, recent theoretical work has pointed out the importance of bipolaron formation for high-temperature superconductivity~\cite{zhang_sous_22} and the experiments conducted in Ref.~\cite{Zhang2022} emphasize the role of bipolarons in the pseudogap formation in the quasi-one-dimensional material (TaSe$_4$)2I.

An prototype microscopic model, capturing the interaction between electrons and optical phonons, is the Holstein model~\cite{Holstein1959}. This model consists of electrons which can move around and interact with local harmonic oscillators. Despite its apparent simplicity, the Holstein model captures essential elements of both single polaron and bipolaron physics. Further, it also exhibits a transition from a charge-density wave (CDW) to a Tomanga-Luttinger liquid (TLL) at half filling in one dimension (see, e.g., Refs.~\cite{bursill_98,creffield_05} for phase diagrams). Due to their importance, the Holstein-polaron model~\cite{fratini_03,barisic02,barisic04,barisic06,barisic08,loos_2006,loos_hohenadler_2006,vidmar10,vidmar11,dorfner_vidmar_15,brockt_17,jansen19,bonca2019,chenyen_20,jansen20,bonca_21}, the Holstein model at finite filling~\cite{bursill_98,creffield_05,jeckelmann99,zhang99,zhao_05,hashimoto_ishihara_17,stolpp2020,weber_2021,jansen21}, the Holstein-Hubbard model~\cite{Fehske_94,Ihle_94,wellein96,bonca00,Bonca_01,werner07,fehske08,golez12a,nocera14,werner_eckstein_15,marsiglio_22}, and other similar systems~\cite{vidmar09,vidmar11c,szabo_21,fetherolf_22} have been subject to intense and ongoing theoretical research.

Recent work has pointed out the effects of phonon hopping on CDW formation~\cite{costa_18}, thermalization properties of a polaron interacting with hard-core bosons~\cite{schoenle_21}, the polaron effective mass, the ground-state optical conductivity, and spectral functions~\cite{Marchand_13,bonca_21,Robinson_22_2}. In this work, we compute the real part of the optical conductivity for the Holstein polaron and bipolaron with dispersive optical phonons at different temperatures. We study weak, intermediate, and strong electron-phonon coupling strengths and work in the intermediate phonon frequency and electron-hopping regime. The combination of different interaction strengths together with a phonon dispersion for the polaron and bipolaron at finite temperatures extends and complements previous studies of the optical conductivity in the Holstein model, e.g., in Refs.~\cite{capone_97,zhang99,fratini_01,schubert_05,fratini_06,wellein98,goodvin11,bonca_21}.

The goal of this work is to characterize how the main features of small and large polarons in a Holstein-type system appear in the optical conductivity. By varying temperature and the phonon bandwidth, we extend the understanding of the basic mechanisms in such theoretical Hamiltonians, which incorporate essential physics induced by electron-phonon interaction. This constitutes an essential step in detecting such features, or lack thereof, in experimental data. Some of our key findings are that a finite phonon bandwidth shifts the center of the optical conductivity to higher or lower frequencies, depending on its shape. This holds for both the polaron and the bound bipolaron. Furthermore, the effect remains even when the temperature is increased enough to significantly alter the spectrum. For small electron-phonon coupling, the polaron-bipolaron spectrum remains similar up to a scaling factor. As the electron-phonon coupling is increased, key distinctions appear, such as a two-phonon emission maximum and a temperature-dependent resonance for the bipolaron, and a stable maximum at the one phonon emission peak for the polaron and only a weak temperature dependence on the spectrum (for the temperatures looked at here).

Computing finite-temperature transport properties of quantum systems can be a challenging task, see Ref.~\cite{bertini_21} for a review of one-dimensional systems. To compute the optical conductivity, we use a density-matrix renormalization group (DMRG) based method. DMRG~\cite{white92,schollwock2005density,schollwock2011density} with matrix-product states (MPS) has already been extensively used to study one-dimensional systems, see, e.g., Refs.~\cite{verstrate2004,daley2004,white04,vidal2004,feiguin2005,barthel2009,stoudenmire2010,feiguin2010,paeckel_2019}. One drawback of DMRG is the unfavorable scaling for systems with a large local Hilbert space, as is the case for the Holstein model. For this reason, several schemes to treat these systems more efficiently have been suggested~\cite{zhang98,jeckelmann98,koehler20,mardazad_21}. In Ref~\cite{jansen20}, we successfully combined the finite-temperature method of purification~\cite{verstrate2004} with tDMRG~\cite{daley2004} and local basis optimization (LBO)~\cite{zhang98} to compute several spectral functions of the Holstein-polaron model.

\begin{figure}[t]
\includegraphics[width=0.99\columnwidth]{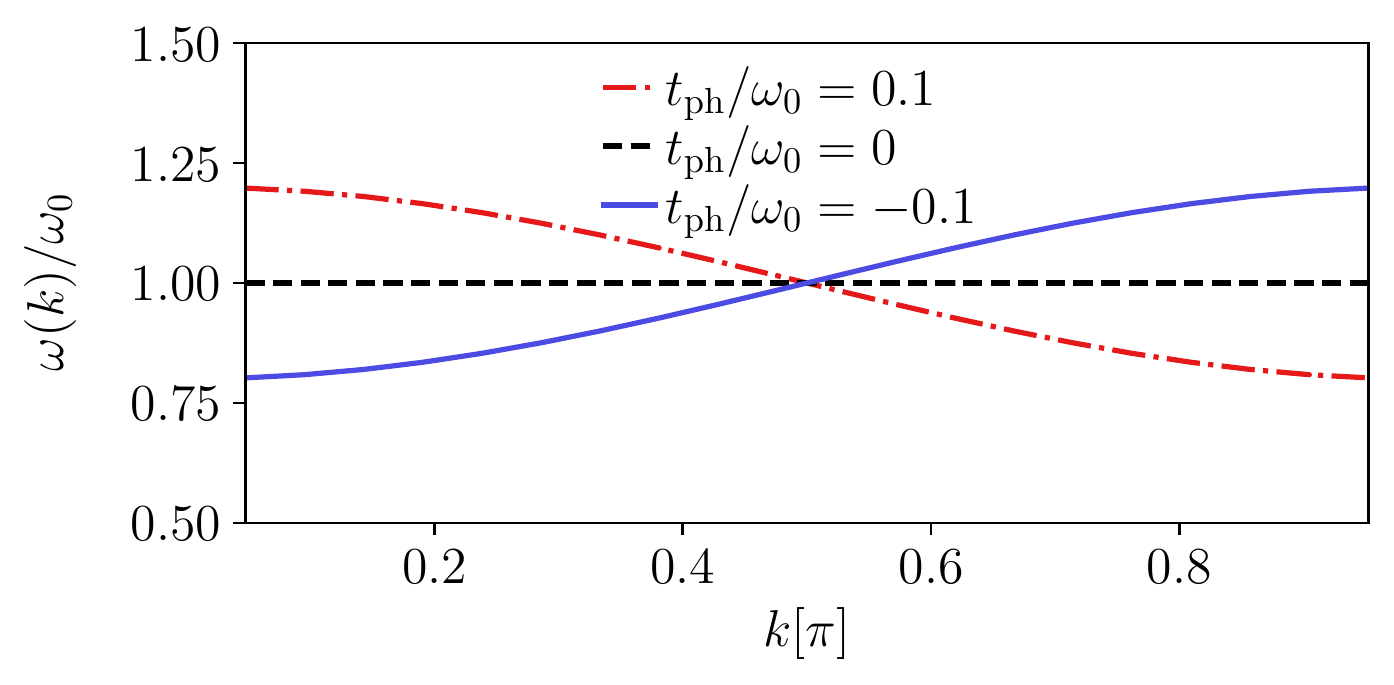}
\caption{Phonon dispersion relation $\omega (k)$ for optical phonons with phonon hopping, see Eq.~\eqref{eq:def_Hph}.  The hopping parameter is $t_{\rm{ph}}$ and the phonons have a finite bandwidth for $t_{\rm{ph}} \neq 0$.}
\label{fig:phonondisp}
\end{figure}

Here, we propose a scheme to combine LBO with the two-site time-dependent variational principle algorithm (TDVP)~\cite{haegeman_11,haegeman_16,secular_2020}. This method is used to conduct the imaginary-time evolution needed to obtain thermal states for weak, intermediate, and strong electron-phonon coupling in the non-adiabatic regime of the Holstein model (the phonon frequency $\omega_0$ is close to the electron tunneling amplitude $t_0$, $t_0\approx\omega_0 $). We then use parallel two-site TDVP~\cite{secular_2020} with LBO (p2TDVP-LBO) to compute the real-time evolution of the current-current correlation function in the ground state for all parameters and at low temperatures in the weak and intermediate coupling regime. For the temperature $T/\omega_0=1$ and a strong electron-phonon coupling,  p2TDVP becomes computationally too demanding and therefore we use the single-site TDVP algorithm (without LBO) instead. We demonstrate that this combination of methods provides access to the real part of the optical conductivity $\sigma^{\prime}(\omega)$. We further add phonon hopping with a hopping amplitude $t_{\rm{ph}}$. This gives rise to the finite phonon bandwidth shown in Fig.~\ref{fig:phonondisp}. We impose $0\leq\abs{t_{\rm{ph}}}/\omega_0 \ll 1$ to model optical phonons. We also compute the real part of the optical conductivity for the bipolaron for weak and intermediate coupling strengths and $t_{\rm{ph}}=0$. Lastly, we analyze the effect of different phonon dispersion relation and a strong electron-phonon coupling on $\sigma^{\prime}(\omega)$ for the bipolaron.

We first test our algorithm by comparing the real part of the optical conductivity with that obtained for the ground state with the Lanczos method~\cite{Lanczos1950,Prelovsek2013} in the weak and intermediate electron-phonon coupling regime. We find a very good agreement. Further, we derive a formula for the real part of the optical conductivity for the Holstein model with a finite phonon bandwidth based on the Born-Oppenheimer Hamiltonian and the Born-Huang ansatz~\cite{Born_27,Born_54}. This formula is consistent with our numerical data for a strong electron-phonon coupling for both the polaron and the bipolaron. In particular, we observe that in this regime, the optical spectrum is similar to that of an asymmetric Gaussian and a downwards phonon dispersion relation (a cosine with a maximum at small momenta, see Fig.~\ref{fig:phonondisp}) shifts the Gaussian to higher frequencies and an upwards phonon dispersion relation (a cosine with a maximum at large momenta, see Fig.~\ref{fig:phonondisp}) shifts it to lower frequencies. Additionally, the upwards phonon dispersion relation leads to less spectral weight at low frequencies for the temperatures studied in this work. These properties hold true for both the polaron and bipolaron.

In the weak and intermediate coupling regime and for the polaron, we see, as previously reported in Ref.~\cite{bonca_21}, that an upwards dispersion relation leads to a continuous spectrum, and a downwards dispersion relation leads to a sequence of distinct peaks for the ground-state optical conductivity.  For a weak electron-phonon coupling strength, the maximum of the spectrum shifts to lower frequencies as the temperature is increased and the different peaks merge. In the intermediate coupling regime, we also observe a significant increase of spectral weight for low frequencies, but the peaks remain well separated for the temperatures investigated here. For the bipolaron, we compare the weak and intermediate coupling for dispersionless phonons. We only detect very small changes in the spectrum for a weak electron-phonon coupling compared to twice the single-polaron curve. At intermediate coupling strengths, however, the spectrum differs significantly and an additional large resonance peak is resolved below $\omega_0$.

This work is structured as follows. In Sec.~\ref{sec:model}, we introduce the Holstein model with dispersive phonons and the optical conductivity, and in Sec.~\ref{sec:BOopc}, we derive a formula for the optical conductivity based on the Born-Oppenheimer Hamiltonian. Section~\ref{sec:methods} introduces the p2TDVP-LBO algorithm. In Sec.~\ref{sec:GSresults}, we compare our Holstein-polaron ground-state results to those obtained with the Lanczos method and analytical calculations. Section~\ref{sec:FTresults} treats the polaron optical conductivity, and Sec.~\ref{sec:Bipolresults} covers the bipolaron, both at finite temperature. Lastly, we conclude in Sec.~\ref{sec:conclusion}.
\section{Model}
\label{sec:model}
 \subsection{The Holstein polaron and bipolaron}
  \label{subsec:holpol}
We consider the Holstein model~\cite{Holstein1959} with one (polaron) and two (bipolaron) electrons. The Hamiltonian is defined as
\begin{equation} \label{eq:def_HolHam}
  \hat H =\hat H_{\rm kin} + \hat H_{\rm ph} + \hat H_{\rm e-ph}\, .
\end{equation}
The model has $L$ sites and we use open boundary conditions. We set $\hbar=1$ throughout this paper.
The first term, the kinetic energy of the electrons,  is
\begin{equation} \label{eq:def_Hkin}
  \hat H_{\rm kin} =- t_0 \sum_{j=1,s={\uparrow, \downarrow}}^{L-1} \left( \hat c_{j,s}^{\dag} \hat c_{j+1,s}^{\phantom{\dag}} + \hat c_{j+1,s}^{\dag} \hat c_{j, s}^{\phantom{\dag}} \right)\, ,
\end{equation}
with $\hat c^{\dag}_{j,s}$ $ (\hat c_{j,s})$ being the electron creation (annihilation) operator on site $j$ with spin $s$ and the hopping amplitude $t_0$. The second term is the phonon energy, consisting of the harmonic  oscillator energy and phonon kinetic energy:
\begin{equation} \label{eq:def_Hph}
    \hat H_{\textrm{ph}}= \omega_0 \sum_{j=1}^L \hat b_j^{\dag}  \hat b_{j}^{\phantom{\dag}} +t_{\rm{ph}} \sum_{j=1}^{L-1} \left( \hat b_j^{\dag} \hat b_{j+1}^{\phantom{\dag}} + \hat b_{j+1}^{\dag} \hat b_j^{\phantom{\dag}} \right)\, , 
  \end{equation}
where $\hat b^{\dag}_j$ $ (\hat b_j)$ creates (annihilates) an optical phonon on site $j$ with frequency $\omega_0$.  The phonons have a hopping amplitude $t_{\rm{ph}}$.  In this work, we always use $t_0/\omega_0=1$. The Hamiltonian in Eq.~\eqref{eq:def_Hph} can be diagonalized by going into quasi-momentum basis by using the Fourier transformation for open boundary conditions  \begin{equation} \label{eq:def_fFTphonon}
\hat{b}_k=\sqrt{\frac{2}{L+1}}\sum\limits_{j=1}^{L}\sin(kj)\hat{b}_j\, , 
  \end{equation} 
  where $k=\pi m_k / (L+1) $ and with $1\leq m_k \leq L$.  Then, 
  \begin{equation} \label{eq:def_Hph_ft}
    \hat H_{\textrm{ph}}=  \sum_{k=\pi / (L+1) }^{\pi L / (L+1) } \omega(k) \hat b_k^{\dag}  \hat b_{k}^{\phantom{\dag}} \, , 
  \end{equation} 
  with $\omega(k)=\omega_0+2t_{\rm{ph}} \cos (k)$.  $\omega(k)$ is the dispersion relation for the optical phonons with a finite bandwidth for $t_{\rm{ph}} \neq 0$.
  In this work, we aim to model optical phonons with zero or small bandwidth and choose $\abs{t_{\rm{ph}}}\ll\omega_0$.  For $t_{\rm{ph}}<0(>0)$, we refer to the phonons as having an upward (downward) dispersion relation. In Fig.~\ref{fig:phonondisp}, we illustrate the phonon dispersion relation for the values of $t_{\rm{ph}}/\omega_0$ used here. 
  
  The final term of the Hamiltonian in Eq.~\eqref{eq:def_HolHam} is the electron-phonon coupling 
\begin{equation} \label{eq:def_Heph}
    \hat H_{\rm e-ph}= \gamma \sum_{j=1}^L\hat n_j \left( \hat b_j^{\dag} + \hat b_{j}^{\phantom{\dag}}  \right) \, ,
  \end{equation}
  with $\hat n_j= \hat c^{\dag}_{j,\uparrow} \hat c_{j,\uparrow}+\hat c^{\dag}_{j,\downarrow} \hat c_{j,\downarrow}$, and the coupling parameter $\gamma$.   Note that since we do not include electron-electron repulsion in this work, we always deal with a bound bipolaron.  Furthermore, we truncate the local phonon-Hilbert space to a finite maximum phonon number $M$. To characterize the transition between a small and a large polaron we introduce the dimensionless parameter
  \begin{equation} \label{eq:def_lambda}
   \lambda=N\frac{\gamma^2}{2t_0\omega_0} \,,
  \end{equation}
  where $N=1$ for the polaron and $N=2$ for the bipolaron.
  
In order to obtain the real part of the optical conductivity, we first define the current-current correlation function as
        \begin{equation} \label{eq:def_corel0}
C_{T}( t)= \expval*{\hat J(t)\hat J(0)}_T \, ,
\end{equation} where $\expval{\hdots}_T$ is the expectation value calculated in the canonical ensemble at temperature $T$.  The current operator $\hat J$ is defined as   \begin{equation} \label{eq:def_curr}
\hat J=it_0 \sum\limits_{i=1,s={\uparrow, \downarrow}}^{L-1}(\hat c_{j,s}^{\dag} \hat c_{j+1,s}^{\phantom{\dag}} - \hat c_{j+1,s}^{\dag} \hat c_{j,s}^{\phantom{\dag}} ) \, .
\end{equation}For calculations in the ground-state $\ket{GS}$, we write $\lim\limits_{T\rightarrow 0}C_{T}( t)=\lim\limits_{T\rightarrow 0}  \expval*{\hat J(t)\hat J(0)}_T = \expval*{\hat J(t)\hat J(0)}{GS}=C_{GS}( t)$. We further define its Fourier transformation as 
        \begin{equation} \label{eq:def_corelFT}
C_{T}( \omega )= \int_{-\infty}^{\infty} e^{i \omega t} f(t)C_{T}(t)  dt \, ,
\end{equation} where $f(t)$ is a damping function ensuring a decay which we either choose to be  $e^{-\eta \abs{t}} $ or $ e^{-\eta t^2} $, leading to a Lorenzian or Gaussian  broadening in frequency space, respectively. In general, we obtain a higher accuracy of the moments of the correlation functions, see Appendix~\ref{sec:app2}, when using Gaussian broadening. Still, we will show data using Lorenzian broadening in Sec.~\ref{sec:GSresults} since this is used by the benchmark data. Furthermore, presenting results with different broadening functions demonstrates the small differences. 

Our goal is to calculate the optical conductivity which can be split into real and imaginary part
                     \begin{equation} \label{eq:def_opc}
      \sigma(\omega)= \sigma^{\prime}(\omega)+i \sigma^{\prime \prime}(\omega) \,  .
        \end{equation}
                The imaginary part is related to the real part through the Kramers-Kronig relation~\cite{stone_goldbart_2009}. 
                The real part of $\sigma (\omega)$ can be extracted from the current-current correlation function via
        \begin{equation} \label{eq:def_opc}
       { \sigma}^{\prime} (\omega)= \frac{1-e^{-\omega /T }}{2\omega}C_T(\omega) \,  .
        \end{equation}
        For the ground state, we have
                \begin{equation} \label{eq:def_opc_T0}
       \sigma^{\prime}_{T\rightarrow0} (\omega)= \frac{1}{2\omega}C_{GS}(\omega) \,  .
        \end{equation}
    \section{Optical conductivity based on the Born-Oppenheimer Hamiltonian}
\label{sec:BOopc}
An analytical formula for the optical conductivity for the Holstein polaron has been obtained in several different contexts and for different limiting cases,  see, e.g., Refs.~\cite{Reik_67,emin_93,mahan_90,fratini_06}. The theoretical results have been shown to fit both numerical,  see, e.g., in Refs.~\cite{schubert_05,fratini_06}, and experimental data, see, e.g., Refs.~\cite{kim_96,yoon_98,hartinger_06,mildner_15}.
In this section, we obtain an expression for the real part of the optical conductivity based on the Born-Oppenheimer (BO) Hamiltonian and the Born-Huang formalism~\cite{Born_27,Born_54}, which captures the influence of the finite phonon bandwidth for both the polaron and bipolaron.  For more details, see Appendix~\ref{sec:app3} and Refs.~\cite{mahan_90,schatz_02,nitzan_06}. We first rewrite Eq.~\eqref{eq:def_HolHam} in terms of the phonon position and momentum operators, using that $\hat b_i=\sqrt{\frac{m \omega_0}{2}}(\hat x_i + \frac{i}{m \omega_0}\hat p_i)$ and $\hat b_i^{\dagger}=\sqrt{\frac{m \omega_0}{2}}(\hat x_i - \frac{i}{m \omega_0}\hat p_i)$. 
The total Hamiltonian has the form
\begin{equation}
\label{eq:def_HofX}
\begin{split}
\hat H & =\hat H_{\textrm{kin}}+\sum\limits_{i=1}^L   \bigg( \gamma \sqrt{\frac{m \omega_0}{2}} \hat n_i(2\hat x_i) \\ & +\frac{m \omega_0^2 }{2}(\hat x_i^2+\frac{1}{m^2 \omega_0^2}\hat p_i^2-\frac{1}{m \omega_0})  \bigg) \\& +t_{\rm{ph}} m \omega_0 \sum\limits_{i=1}^{L-1}(\hat x_i \hat x_{i+1}+\frac{1}{m^2 \omega^2_0}\hat p_{i}\hat p_{i+1}) \, .
\end{split}
\end{equation}

\begin{figure}
\includegraphics[width=0.99\columnwidth]{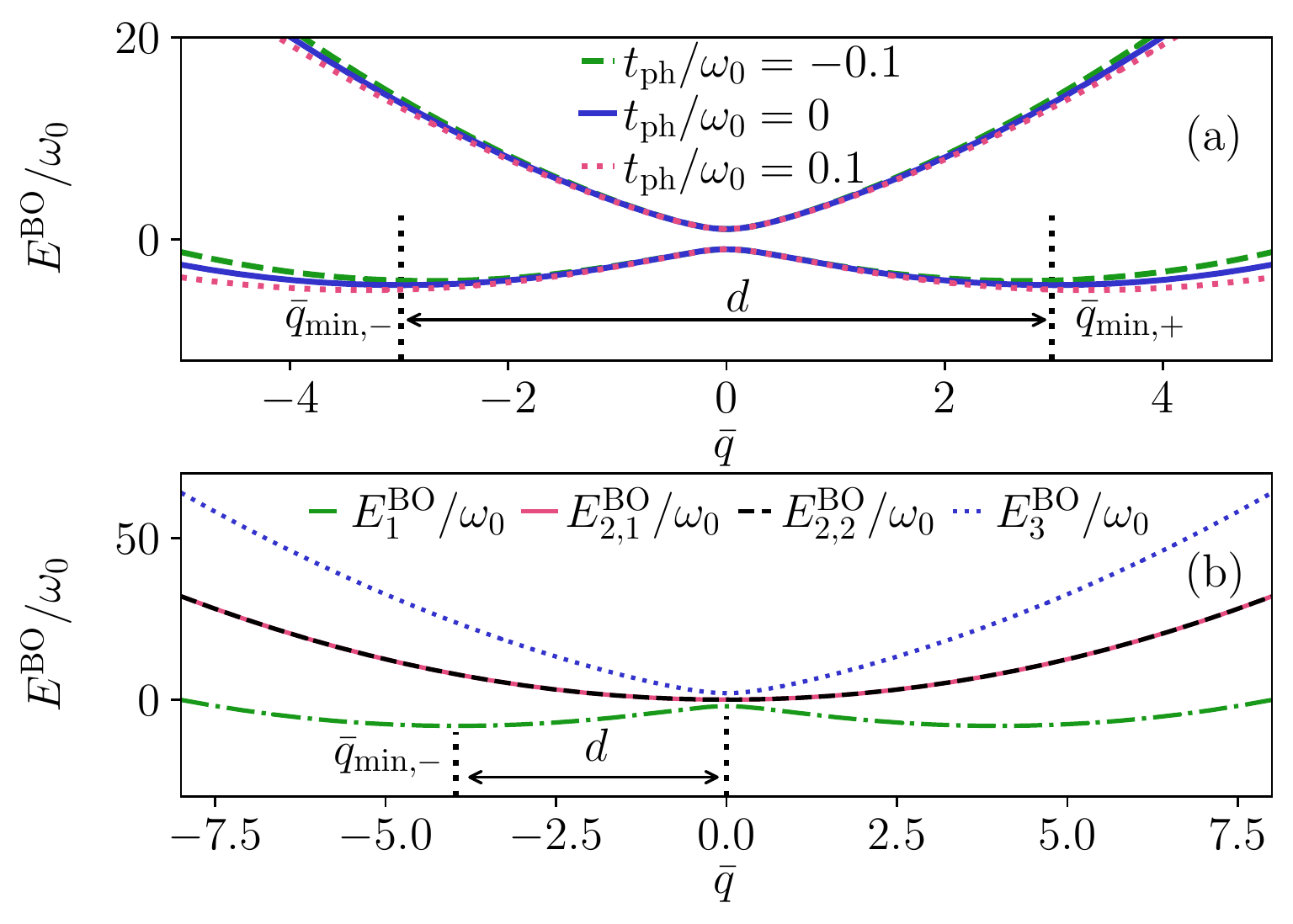}
\caption{Born-Oppenheimer surfaces for the polaron and bipolaron in the Holstein dimer. (a) The two polaron Born-Oppenheimer surfaces with different phonon dispersion $t_{\rm{ph}}/\omega_0$. We use $\gamma/\omega_0=3$ and $t_0/\omega_0=1$. (b) The four Born-Oppenheimer surfaces for the Holstein-dimer bipolaron with $\gamma/\omega_0=2$, $t_{\rm{ph}}/\omega_0=0$ and $t_0/\omega_0=1$. }%
\label{fig:BOsurf_comb}
\end{figure}

\begin{figure}[t]            
\includegraphics[width=0.99\columnwidth]{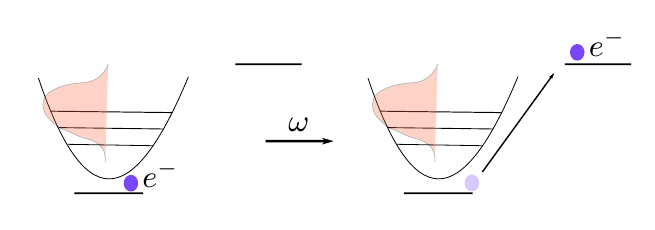}
    \caption{Illustration of the Franck-Condon excitation, which is the process leading to Eq.~\eqref{eq:def_analyt_formSC}. The electron is localized on one site  in a potential due to the coupling to the phonons (illustrated by the shaded Gauss shape in the harmonic oscillator).   Shining light with frequency $\omega=2E_p$ will lead to the electron escaping the potential generated by the phonons and moving to the next site.  During this process,  the phonon configuration remains unchanged. This corresponds to a diagonal transition from the lower to the upper Born-Oppenheimer surface in Fig.~\ref{fig:BOsurf_comb}(a).}  
    \label{fig:pic}        
\end{figure}

We will now consider the Holstein dimer, i.e., $L=2$,  and go to the relative and center-of-mass coordinates with $\hat q=\frac{1}{\sqrt{2}}(\hat x_1-\hat x_2)$ and $\hat Q=\frac{1}{\sqrt{2}}(\hat x_1+\hat x_2)$.  Using the fact that the center-of-mass coordinates $\hat Q$ are independent of the rest of the system for constant electron density, going in the adiabatic limit of slow phonons (sending all momenta to zero) and sending $\hat q\rightarrow  q $, we can write the Born-Oppenheimer Hamiltonian as \begin{equation} \label{eq:def_BOHam}
\hat H_{\textrm{BO}}= \hat H_{\textrm{kin}} +  \gamma[ \bar{q}(n_1-n_2)]+  \frac{1}{2 } \bar{q}^2(\omega_0-t_{\textrm{ph}}) \, .
\end{equation}
Here, we have defined $\bar{q}=q\sqrt{\frac{1}{m\omega_0}}$. $\hat H_{\textrm{BO}}$ is now a $2\cross 2$ and a $4\cross 4$ matrix for the polaron and bipolaron, respectively, see, e.g., Refs.~\cite{firsov_01,McKemmish_15}. These Hamiltonians can be diagonalized and one obtains two or four Born-Oppenheimer surfaces, where two are degenerate in the latter case.  We label the surfaces $E^{BO}_1$ and $E^{BO}_2$ for the polaron and $E^{BO}_1,E^{BO}_{2,1}=E^{BO}_{2,2}$, and $E^{BO}_3$ for the bipolaron.  The polaron and bipolaron surfaces are shown in Figs.~\ref{fig:BOsurf_comb}(a) and \ref{fig:BOsurf_comb}(b) for some of the parameters used later in this work. The lowest surface has two minima which differ by a sign. We refer to these as  $ \bar{q}_{\textrm{min},\pm}$. The current operator from Eq.~\eqref{eq:def_curr} can connect the states in the lowest and first excited surface while leaving the phonon configuration invariant. This process, also known as a Franck-Condon excitation, is illustrated in Fig.~\ref{fig:pic} and corresponds to a vertical transition into the next surface at a fixed $\bar q$. When $\abs{\bar{q}_{\textrm{min},\pm}}$ is large, this can be approximated as a transition between two harmonic oscillator potentials separated by a distance $d$ and with an energy shift $\Delta_E$. This leads to 
 \begin{equation} \label{eq:def_analyt_formSC}
 \begin{split}
     \sigma_{SC} (\omega)=N\frac{1-e^{-\omega /T }}{\omega}t_0^2 \;\sqrt{ \frac{4 \pi}{d^2 \coth (\frac{ \tilde{\omega}_0}{2T})\tilde{\omega}_0^2  }} 
     \\ \cross e^{-(\Delta_E+\frac{d^2\tilde{\omega}_0}{2}-\omega )^2/( d^2 \tilde{\omega}_0^2 \coth (\frac{ \tilde{\omega}_0 }{ 2T}))} \, .
     \end{split}
        \end{equation}
        Here, we set $\tilde{\omega}_0=\omega_0-t_{\rm{ph}}$.
        For the Holstein polaron, $d=2\bar{q}_{\textrm{min}} $ and $\Delta_E=0$. For the Holstein bipolaron, we have $d=\bar{q}_{\textrm{min}}$ and $\Delta_E=E^{BO}_1(\bar{q}_{\textrm{min}})$, as illustrated in Fig.~\ref{fig:BOsurf_comb}. Note that Eq.~\eqref{eq:def_analyt_formSC} is an asymmetric Gaussian around $\Delta_E+d^2\tilde{\omega}_0 $ which is the energy difference between the lowest and 
        the first excited Born-Oppenheimer surface.  For the polaron,  everything can be solved analytically and one gets $ \bar{q}_{\textrm{min},\pm}=\pm \frac{\sqrt{\gamma^4-\tilde{\omega}_0^2{t}_0^2}}{\tilde{\omega}_0\gamma}$. For the bipolaron,  we obtain the Born-Oppenheimer surfaces numerically. In the case of the polaron, with $T\rightarrow 0 $, $\gamma \gg t_0$ and $t_{\rm{ph}}=0$, we have $ \bar{q}_{\textrm{min},\pm}= \pm \frac{\gamma}{\omega_0}$ and 
        \begin{equation} \label{eq:def_analyt_formSC_lim}
 \begin{split}
\sigma_{SC} (\omega)& = \frac{t_0^2}{\omega}\;\sqrt{ \frac{ \pi}{ \frac{\gamma^2}{ \omega_0} \omega_0  }}
      e^{(2 \frac{\gamma^2}{\omega_0}-\omega )^2/( 4 \frac{\gamma^2}{\omega_0} \omega_0 )}\\  &=\frac{t_0^2}{\omega} \; \sqrt{ \frac{ \pi}{ E_P \omega_0  }} 
      e^{(2 E_P-\omega )^2/( 4 E_P \omega_0 )}\,  , 
     \end{split}
        \end{equation}
        which is the well-known result of an asymmetric Gaussian around twice the polaron binding energy $E_P=\gamma^2/\omega_0$. 

\section{Methods}
\label{sec:methods}
There have already been many applications utilizing the matrix-product state formalism for finite temperature calculations, see, e.g., Refs.~\cite{verstrate2004,feiguin2005,white2009,barthel2009,
stoudenmire2010,feiguin2010,binder_15}. In this work, we use purification~\cite{verstrate2004,feiguin2005,barthel2009,feiguin2010,barthel_16}.  In this scheme, one simulates the thermal density matrix of the system by introducing ancillary sites,  thus doubling it. The density matrix at the desired temperature $T$ can then be obtained by starting with the infinite-temperature state, imaginary time evolving it to the inverse temperature $1/(2T)$ and then tracing out the ancillary sites. In other words, one obtains the thermal state living in the physical and ancillary Hilbert spaces,  $H _{P}$ and $H _{A}$, respectively,  i.e., $\ket*{\psi_T}\in H _{P}\otimes H_{A} $. The thermal expectation value of an observable $\hat O$ acting on a state in $ H _{P}$ is given by $\expval*{\hat O}_T=\expval*{\hat O}{\psi_T}$. The state $\ket{\psi_{\infty}}$ can often be obtained either analytically or by finding the ground state of a suitable Hamiltonian~\cite{nocera_16},  and $\ket*{\psi_{T}}$ can be calculated by conducting imaginary time evolution on $\ket{\psi_{\infty}}$. Note that even though the infinite-temperature state is artificial for our implementation of the Holstein model due to the finite phonon truncation $M$, this method still captures the correct low-temperature physics, see Ref.~\cite{jansen20}. There, a detailed description of how to obtain $\ket*{\psi_{\infty}}$ for the Holstein polaron model is also given. 

To obtain the time-dependent correlation functions from Eq.~\eqref{eq:def_corel0}, one needs to carry out a real-time evolution after the imaginary-time evolution. The real-time evolution is, however,  not carried out on   $\ket*{\psi_{T}}$, but rather on the state $\ket*{\phi_{T}}$, which is obtained by acting on the thermal states $\ket*{\psi_{T}}$ with the current operator $\hat J$,  $\ket*{\phi_{T}}=\hat J \ket*{\psi_{T}}$. The real-time evolution of  $\ket*{\phi_{T}}$ is the computationally most costly part of our procedure. 

There are several ways to time evolve matrix-product states, see Ref.~\cite{paeckel_2019} for a review, including the time-evolution block decimation (TEBD) algorithm, see Ref.~\cite{vidal2004},  which also has a parallel implementation~\cite{Urbanek_16}, and the time-dependent variational principle (TDVP) algorithm, see, e.g., Refs.~\cite{haegeman_11,haegeman_16,kloss_18,kloss_19,yang_20,Xu_21}.   In this work, we use the single-site TDVP,  the two-site TDVP, and the parallel two-site TDVP (p2TDVP) algorithms introduced in Ref.~\cite{secular_2020}. We tailor the latter two specifically for electron-phonon systems with local basis optimization (LBO)~\cite{zhang98}. Since the single-site and two-site TDVP are standard in the literature~\cite{haegeman_11,haegeman_16,paeckel_2019}, we will, in Sec.~\ref{subsec:ptdvp},  only review the key ideas of parallelization of the two-site TDVP,  and focus on how we incorporate LBO into the algorithm. We emphasize that the main idea of the parallelization of TDVP is based on the parallel ground-state DMRG (pDMRG) method introduced by Stoudenmire and White in Ref.~\cite{stoudenmire2013} and that both the pDMRG and the pTDVP algorithms are described comprehensively in Refs.~\cite{stoudenmire2013} and~\cite{secular_2020}, respectively. 

    \subsection{Parallel two-site time-dependent variational principle}
\label{subsec:ptdvp}
\begin{figure}[t]
\includegraphics[width=0.99\columnwidth]{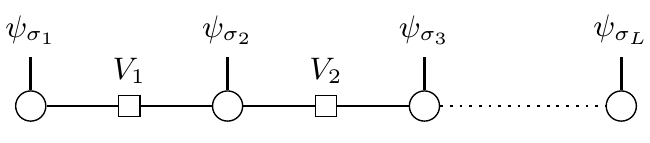}
\caption{Matrix-product state in the inverse canonical form.} 
\label{fig:Canform}
\end{figure}

\begin{figure}[t]
\includegraphics[width=0.99\columnwidth]{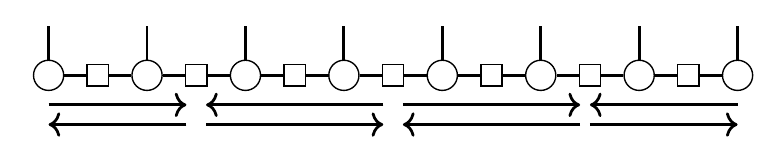}
    \caption{Sweeping scheme of the pTDVP and pDMRG algorithms. One process starts sweeping either to the left or to the right as indicated by the arrows.  At each contact point, the local tensors are shared and one process performs the sweeping step at the boundary. The resulting local tensor is then transported to its corresponding process and the sweeping is then continued in the opposite direction. In this example, the program would run on 4 processes.}
\label{fig:parsweep_pic}
\end{figure}

For both pDMRG and p2TDVP, the first step is to bring the matrix-product state into the inverse canonical-gauge form
 \begin{equation}
 \label{eq:invcanform}
\ket{\psi}=\sum\limits_{\vec{\sigma}} \psi_{\sigma_1} V_1 \psi_{\sigma_2} V_2 \hdots V_{L-1} \psi_{\sigma_L} \ket{\vec{\sigma}} \,  .
\end{equation}
Here, for a clearer notation, we will not write the bond indices unless they are specifically needed, and $\ket{\vec{\sigma}}=\ket{\sigma_1, \sigma_2, \hdots, \sigma_L}$, with $\sigma_l$ being the physical state on site $l$. The state is depicted in Fig.~\ref{fig:Canform}. For the Holstein polaron model ,$\ket*{\sigma_l}$ consists of the local electron and phonon state $\ket*{\sigma_l}=\ket*{n_e,n_{ph}}$, with $n_e=\{0,1\} $ and $n_{ph}=\{0,\hdots, M\} $.
Equation~\eqref{eq:invcanform} can easily be obtained from the state in the canonical gauge~\cite{vidal2003,vidal2004}
\begin{equation}
 \label{eq:canform}
\ket{\psi}=\sum\limits_{\vec{\sigma}} \Gamma_{\sigma_1} \Lambda_1 \Gamma_{\sigma_2} \Lambda_2 \hdots \Lambda_{L-1} \Gamma_{\sigma_L} \ket{\vec{\sigma}} \,  ,
\end{equation}
by inserting $V_j \Lambda_j =\mathbb{1}$ and defining $\psi_{\sigma_{j}}=\Lambda_{j-1}\Gamma_{\sigma_j} \Lambda_j$. The key ingredient of pDMRG is to partition the wave function from Eq.~\eqref{eq:invcanform} onto several processes which each do their DMRG sweep and only communicate when the respective processes reach their corresponding boundaries. To make this as efficient as possible, one lets two processes start sweeping in different directions and communicate upon return. The scheme is sketched in Fig.~\ref{fig:parsweep_pic}.

The serial TDVP algorithm can be formulated similarly as DMRG, but by solving local time-dependent equations.
For each update,  here illustrated for the left-to-right sweep,  one proceeds by first computing $\theta_{\alpha \beta }=\psi_{\sigma_i} V_i \psi_{\sigma_{i+1}} $, where the index $\alpha $ $(\beta)$ contains the local degree of freedom $\sigma_i$ $(\sigma_{i+1})$ and the left (right) bond $m_l$ $(m_r)$, i.e., $\alpha=(\sigma_i, m_l)$ and $\beta=(\sigma_{i+1}, m_r)$. The tensor $\theta_{\alpha\beta }$ is updated by solving the equation
\begin{equation}
 \label{eq:eq_to_solve}
\dot{\theta}_{\alpha \beta}(t)=\sum\limits_{\alpha^{\prime},  \beta^{\prime}}-i\frac{ d t}{2} H^{\textrm{eff}}_{\alpha \beta  \alpha^{\prime}\beta^{\prime} } \theta_{\alpha^{\prime}  \beta^{\prime} } (t)\, ,
\end{equation} 
where $H^{\textrm{eff}}_{\alpha \beta  \alpha^{\prime}\beta^{\prime} } $ is the effective two-site Hamiltonian, calculated by contracting the rest of the matrix-product state with the matrix-product operator Hamiltonian, see e.g., Refs.~\cite{schollwock2011density,paeckel_2019,secular_2020} for details.  Equation~\eqref{eq:eq_to_solve} has the solution \begin{equation}
 \label{eq:eq_solved}
\theta_{\alpha\beta }\Bigl(t+\frac{dt}{2}\Bigr)=\sum\limits_{\alpha^{\prime},  \beta^{\prime}}e^{-i\frac{ dt}{2} H^{\textrm{eff}}_{\alpha \beta  \alpha^{\prime}\beta^{\prime} }}  \theta_{\alpha^{\prime} \beta^{\prime} }(t)\,  .
\end{equation}

Consecutively, a singular value decomposition (SVD) is performed on the updated $\theta_{\alpha\beta }=A_{m_l \sigma_i} \Lambda_i B_{\sigma_{i+1}m_r}  $ and $V_i  \Lambda_i $ is inserted.  We then write $\psi_{\sigma_i}=A_{m_l\sigma_i} \Lambda_i$ and $\psi_{\sigma_{i+1}}= \Lambda_{i} B_{\sigma_{i+1}m_r}$.  For the discussion of the local basis optimization in the algorithm, we refer to solving Eq.~\eqref{eq:eq_to_solve} as step one and the following SVD as step two. 
Lastly, the effective one-site Hamiltonian $H^{\textrm{eff}}_{\gamma  \gamma^{\prime}}$ for site $j+1$ is computed and $\psi_{\gamma}$ is evolved backwards in time by solving
\begin{equation}
 \label{eq:eq_to_solve2}
\psi_{ \gamma }(t)=\sum\limits_{ \gamma^{\prime}}e^{+i \frac{dt}{2} H^{\textrm{eff}}_{\gamma  \gamma^{\prime}}} \psi_{\gamma^{\prime}}\Bigl(t+\frac{dt}{2}\Bigr)\,  ,
\end{equation} 
where we wrote all indices of $\psi_{\sigma_{i+1}} $ as $\gamma$ for brevity.
The same process is then repeated for sites $i+1$ and $i+2$. 
Similar to pDMRG, one can partition the system for p2TDVP. We illustrate this in Fig.~\ref{fig:parsweep_pic}.This is done under the assumption that the inverse canonical form is approximately preserved for small $dt$. Thereafter, one assigns each partition to an individual process together with the corresponding effective Hamiltonian.  
For the truncation in the SVDs, we discard all singular values such that 
\begin{equation}
 \label{eq:bond_cutoff}
\sum\limits_{  \textrm{discarded } \: \eta} s_{\eta}^2/(\sum\limits_{\textrm{all} \:  \eta } s_{\eta}^2)<\epsilon_{\rm bond}\,  .
\end{equation}  
    \subsection{Local basis optimization}
\label{subsec:lboalg}
We now discuss how we incorporate the local basis optimization in the 2TDVP algorithm. This procedure is independent of whether the 2TDVP is run serially or in parallel. The idea of LBO~\cite{zhang98} is to find a so-called optimal basis in which the state can be represented efficiently and truncated with a negligible error. LBO has already been used in many applications, see, e.g., Refs.~\cite{zhang99,Guo2012,brockt_dorfner_15,brockt_17,stolpp2020,jansen20,jansen21}.  To obtain the transformation matrices into an optimal basis, one first computes the reduced density matrix from, in our algorithm, the two-site tensor 
\begin{equation}
\label{eq:def_redm}
\rho_{ \sigma_{i}  \sigma_i^{\prime}}=(\theta \theta^{\dagger} )_{ \sigma_{i}    \sigma_i^{\prime}}\,  ,
\end{equation} where all remaining indices have been contracted.
$\rho_{ \sigma_{i} \sigma_i^{\prime}}$ is then diagonalized
\begin{equation}
\label{eq:def_eddmdiag}
\rho=R^{\dagger} WR \,  .
\end{equation}
The matrix $R$ transforms the site from the physical basis $\sigma_i$ to the optimal basis $\tilde{\sigma}_i$ and $W$ is a diagonal matrix with the values $w_{\alpha}$. The truncation of the optimal basis from the local dimension $d$ into the dimension $d_{\rm LBO}$ is done based on the magnitude of the eigenvalues $w_{\alpha}$, which, in many cases, decay exponentially, see, e.g., Refs.~\cite{dorfner_vidmar_15,jansen21}. When diagonalizing the reduced density matrices in Eq.~\eqref{eq:def_redm} to obtain the optimal basis,  the smallest eigenvalues $w_{\alpha}$ are discarded such that the truncation error is below a threshold: \begin{equation}
\label{eq:def_lbocut}
\sum\limits_{  \textrm{discarded} \: \eta } w_{\eta}/(\sum\limits_{\textrm{all} \: \eta} w_{\eta})<\epsilon_{\rm LBO}\,  .
\end{equation}

 There are two costly operations where a transformation of the physical index to an optimal basis can be beneficial. The most costly operation of the algorithm is the contraction of the effective Hamiltonian $H^{\textrm{eff}}_{\alpha \beta  \alpha^{\prime}\beta^{\prime} }$ with the local tensor $\theta_{\alpha^{\prime} \beta^{\prime} }$ needed to solve Eq.~\eqref{eq:eq_to_solve} in step one. This has a cost of order $\mathcal{O}(d^3m^2D^2+d^2mD^3)$, where $m$ is the bond dimension of the Hamiltonian and $D$ the bond dimension of the state. We solve Eq.~\eqref{eq:eq_to_solve} with a fourth-order Runge-Kutta algorithm (RK4), see, e.g.,  Ref.~\cite{feiguin_05} for an application of RK in the context of DMRG.

The RK4 algorithm consists of approximating $\theta_{\alpha\beta }(t+\frac{\delta t}{2})$ as 
\begin{equation}
\label{eq:def_rk4}
\begin{split}
\theta_{\alpha\beta }\Bigl(t+& \frac{dt}{2}\Bigr)=\theta_{\alpha\beta }(t) \\ &+ \frac{-id t}{2}\frac{1}{6}(k^1_{\alpha\beta}
+2k^2_{\alpha\beta }+2k^3_{\alpha \beta }+k^4_{\alpha \beta})\,  ,
\end{split}
\end{equation}
where for each $k^j$, we must contract the effective Hamiltonian with a two-site tensor.
To incorporate LBO into the algorithm (RK4-LBO), we first obtain the transformation matrices into the local optimal basis states $\tilde{\sigma}_i$ and  $\tilde{\sigma}_{i+1}$ from  $\theta_{\alpha \beta }$ as previously explained. We then transform the two sites of the effective Hamiltonian and $\theta_{\alpha \beta }$ into this new basis, $H^{\textrm{eff}}_{\alpha \beta  \tilde{\alpha} \tilde{\beta}} $ and $\theta_{\tilde{\alpha} \tilde{\beta} }$,  where  $\tilde{\alpha}=(\tilde{\sigma}_i, m_l)$. Then, we compute
\begin{equation}
\label{eq:def_k1}
k^1_{\alpha\beta}=\sum\limits_{\tilde{\alpha}^{\prime} \tilde{\beta}^{\prime} } H^{\textrm{eff}}_{\alpha \beta  \tilde{\alpha}^{\prime} \tilde{\beta}^{\prime} } \theta_{\tilde{\alpha}^{\prime} \tilde{\beta}^{\prime} }(t)\,  .
\end{equation}
We proceed by calculating
\begin{equation}
\label{eq:def_k2}
k^2_{\alpha \beta}=\sum\limits_{\tilde{\alpha}^{\prime} \tilde{\beta}^{\prime}}  H^{\textrm{eff}}_{\alpha \beta \tilde{\alpha}^{\prime} \tilde{\beta}^{\prime}  }  \Bigl(\theta(t)+\frac{-id t}{2}\frac{1}{2}k^1 \Bigr)_{\tilde{\alpha}^{\prime} \tilde{\beta}^{\prime} }\,  ,
\end{equation}
but now $\tilde{\alpha}$ and $\tilde{\beta}$ are the optimal basis states extracted from the tensor $(\theta(t)+\frac{-id t}{2}\frac{1}{2}k^1 )_{{\alpha} {\beta}  }$. A similar procedure is used to obtain $k^3$ and $k^4$.  One first adds the needed tensors and then obtains the new optimal basis states before contracting with the environment.
Note that it is not ensured that one gains a computational benefit from the algorithm. As mentioned earlier, one must first create the reduced density matrices at a cost of order $\mathcal{O}(D^2d^3)$, diagonalize them at a cost of order $\mathcal{O}(d^3)$, apply the transformation matrices at a cost of order $\mathcal{O}(D^2d^2d_{\rm LBO})$, and add the tensors at a cost of order $\mathcal{O}(D^2d^2)$ to get a cost for the contraction of order $\mathcal{O}(d^2d_{\rm LBO}m^2D^2+d^2mD^3)$.  In practice, however, we find this scheme to lead to a computational speed-up in many situations, compared to only implementing step two. This is also due to the additional benefit of the optimal basis on the many sub-leading tensor contractions.

A common approach to solve Eq.~\eqref{eq:eq_to_solve} is to use a Krylov method.  While the RK4 algorithm is non-unitary we also compare some of our data to that obtained using a Krylov solver without LBO and the tDMRG-LBO time-evolution algorithm used in Ref.~\cite{jansen20}. The error appears to be negligible in the cases studied here. A comparison and discussion of the different algorithms are contained in Appendix~\ref{sec:app1}. Alternatively, one could also apply the local basis transformation when using a Krylov solver,  thus needing to find the optimal basis for each Krylov vector. In our tests,  this seems to require more optimal basis states than the RK4 scheme. This is addressed in Appendix~\ref{sec:app1} as well. 

To implement the LBO before the SVD in step two, all one needs to do is to get the local basis of $\theta_{\alpha \beta }(t+\frac{dt }{2})$, then transform the tensor into $\theta_{\tilde{\alpha} \tilde{\beta} }(t+\frac{dt }{2})$, and carry out the singular value decomposition before transforming the tensor back to the bare basis. This step has the cost $\mathcal{O}(d^3+D^2d^3)$, but the cost of the SVD becomes of the order $\mathcal{O}(D^3d_{\rm LBO}^3)$. 

In our experience, only implementing step two alone is sufficient to gain a significant computational speed up. This can easily be incorporated into ground-state DMRG algorithms as well.  
A discussion of the number of local states needed is presented in Appendix~\ref{sec:app1}. 

For the imaginary-time evolution used to obtain all the thermal states in this work, we only implement step two and use a Krylov solver for the local differential equations in Eq.~\eqref{eq:eq_to_solve}. Our reason is that the imaginary-time evolution is less costly, and RK4 is unstable in our calculations for all but very small imaginary time steps $d \tau$. This computation is also run serially. 
To ensure that the low bond dimension of the initial states does not lead to a significant projection error, we compared the states to those obtained using tDMRG for the polaron. For both the polaron and bipolaron, we also compared the states where we imaginary-time evolved using a higher bond dimension through the sub-space expansion introduced in Ref.~\cite{yang_20}.  If $\ket*{\psi_A}$ and $\ket*{\psi_B}$ are states obtained with different imaginary time-evolution schemes, we had $\abs*{1-\ip{\psi_A}{\psi_B}}$ of the order of  $\mathcal{O}(10^{-6})$ in all cases.

For the real-time evolution, we implement both steps one and two and refer to the algorithm as p2TDVP-LBO. Note that when carrying out the real-time evolution, we simultaneously time-evolve the part of the state in the ancillary space back in time to get a slower increase in the bond dimension~\cite{karrasch2012,barthel2012,barthel2013,karrasch2013,kennes2016}. While not an optimal procedure, see Ref.~\cite{hauschild2018},  this can easily be incorporated into the algorithm.  We carry out the real-time evolution until $t_{\rm max }\omega_0$ and add  $4\cdot t_{\rm max }\omega_0$ zeros to the signal for better resolution (zero padding) before the Fourier transformation. Alternatively, one can use linear prediction, see, e.g., Refs.~\cite{vaidyanathan_08,white_affleck2008,barthel2009}, for a higher frequency resolution. We confirmed that both methods give consistent results for our data. Recently, another method for reaching longer times has been proposed~\cite{tian_21}. 

For the real-time evolution, we use a time step $d t/ \omega_0=0.01$ and for the imaginary-time evolution $d \tau/ \omega_0=0.1$. When real-time evolving at high temperatures (here $T/\omega_0=1$) and in the large-coupling regime, we take advantage of the fact that the thermal states have a relatively large bond dimension (which increases further when we act with $\hat J$ on the state) and use a serial single-site TDVP algorithm.  This is done since the bond dimension becomes unfeasible for our computational resources otherwise. Furthermore, this approach does not require the implementation of LBO for our applications, since the algorithm scales better with the local Hilbert space dimension. In this case, $\epsilon_{\textrm{bond},J}$ indicates the cutoff at which we truncate after applying the current operator to the thermal state. Here, convergence is monitored by changing $\epsilon_{\textrm{bond},J}$,  even though the time-evolution scheme itself has a constant bond dimension.  To make sure that small bond dimensions on some bonds do not lead to errors, we also compared the results to those obtained with a larger bond dimension by enforcing a minimum bond dimension when applying $\hat J$.

To further verify our results, we carefully compare the moments of the time-dependent correlation functions to their finite-temperature expectation values, see Appendix~\ref{sec:app2} for details. To test our algorithm,  we also compare our ground-state optical conductivity to data obtained with the Lanczos method for weak and intermediate electron-phonon coupling. Furthermore, we check that our finite-temperature polaron data in the weak and intermediate electron-phonon coupling regime is consistent with those obtained with a different time-evolution method whenever possible, namely the tDMRG-LBO method used in Ref.~\cite{jansen20}. Additionally, we monitor the $f$-sum rule in the large electron-phonon coupling regime. This is also contained in Appendix~\ref{sec:app2}. 
All calculations are done with the ITensor software library~\cite{itensor}. 

\begin{figure}[t]
\includegraphics[width=0.99\columnwidth]{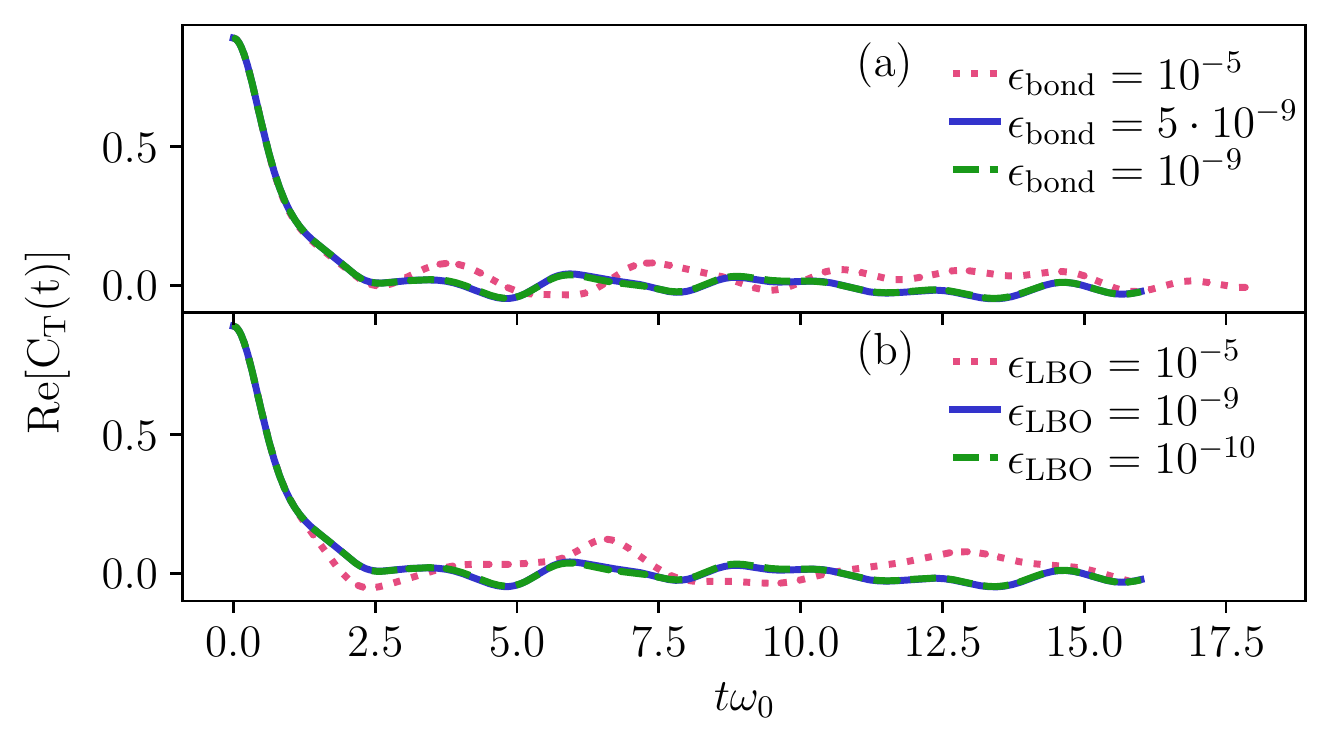}
\caption{(a) Real part of the current-current correlation function from Eq.~\eqref{eq:def_corel0} for $L=20$, $M=20$, $\gamma/\omega_0=\sqrt{2}$, and $t_{\rm{ph}}/\omega_0=-0.1$, computed in the ground state ($T/\omega_0=0$). We use a fixed $\epsilon_{\rm LBO}=10^{-9}$ and different $\epsilon_{\rm bond}$. (b) Same as in (a) but with a fixed $\epsilon_{\rm bond}=10^{-9}$ and different $\epsilon_{\rm LBO}$. The calculation is done with p2TDVP-LBO and is distributed onto four processes.} 
\label{fig:convplot}
\end{figure}

We now demonstrate how we control the convergence of the p2TDVP-LBO time evolution with respect to $\epsilon_{\textrm{bond}}$ and $\epsilon_{\textrm{LBO}}$.  This is illustrated in Fig.~\ref{fig:convplot} for the real-time current-current correlation function for $\gamma/\omega_0=\sqrt{2}$ and $t_{\rm{ph}}/\omega_0=-0.1$.  Figure~\ref{fig:convplot}(a) shows $\textrm{Re}[C_T(t)]$ for a fixed $\epsilon_{\textrm{LBO}}$ and different $\epsilon_{\textrm{bond}}$. The data illustrate that when $\epsilon_{\textrm{bond}}$ is chosen sufficiently small, the correlation functions become independent thereof and the curves are converged for our purposes.  In Fig.~\ref{fig:convplot}(b), we show a similar plot but for a fixed $\epsilon_{\textrm{bond}}$.  As can be seen, both $\epsilon_{\textrm{bond}}$ and $\epsilon_{\textrm{LBO}}$ must be chosen carefully.  We note that during this work, we have observed that, in some cases, one can obtain similar optical conductivities with the same key features even when the correlations functions still depend noticeably on cutoffs. One could control convergence by only monitoring the quality of the optical conductivity. We work with the stricter criteria of monitoring both the correlation functions and the optical conductivity as elaborated on above. Note that in some cases, small deviations could not be avoided with our computational resources.

\section{Ground-state results}
\label{sec:GSresults}
In this section, we present the results for the real part of the optical conductivity obtained with p2TDVP-LBO for the ground state of the Holstein polaron.  The ground state itself is obtained with ground-state DMRG. Our results are compared to calculations with the Lanczos method for weak and intermediate electron-phonon coupling. For strong coupling, we compare to Eq.~\eqref{eq:def_analyt_formSC}.  

 \subsection{Ground-state results in the weak- and intermediate-coupling regimes}
  \label{subsec:opcGSWIC}
     We first look at the optical conductivity in the ground state for different electron-phonon coupling strengths and different phonon dispersion relations. The effects of the dispersion on the optical conductivity were thoroughly investigated in Ref.~\cite{bonca_21}, however,  we also describe the behavior here for self-consistency and to guide the discussion of the results at finite temperature. 
    
    We verify the correctness of our results by comparing them to the optical conductivity obtained with the Lanczos method. This method uses a constructed variational Hilbert space and allows for 28 phonons at the initial site of the electron, but fewer and fewer at the sites further away, see Ref.~\cite{bonca_21} for details. The Lanczos method has the advantage of computing the optical conductivity directly in frequency space and can easily extract the Drude weight. However, it has more difficulties in including many phonons  and larger systems than the DMRG method.  Lastly, the DMRG has an (in principle) straightforward extension to finite densities. In a setup where the Lanczos method cannot utilize the variational Hilbert space approach, the treatable system sizes will also be limited.
  
\begin{figure}[t]
 \includegraphics[width=0.99\columnwidth, height=10cm]{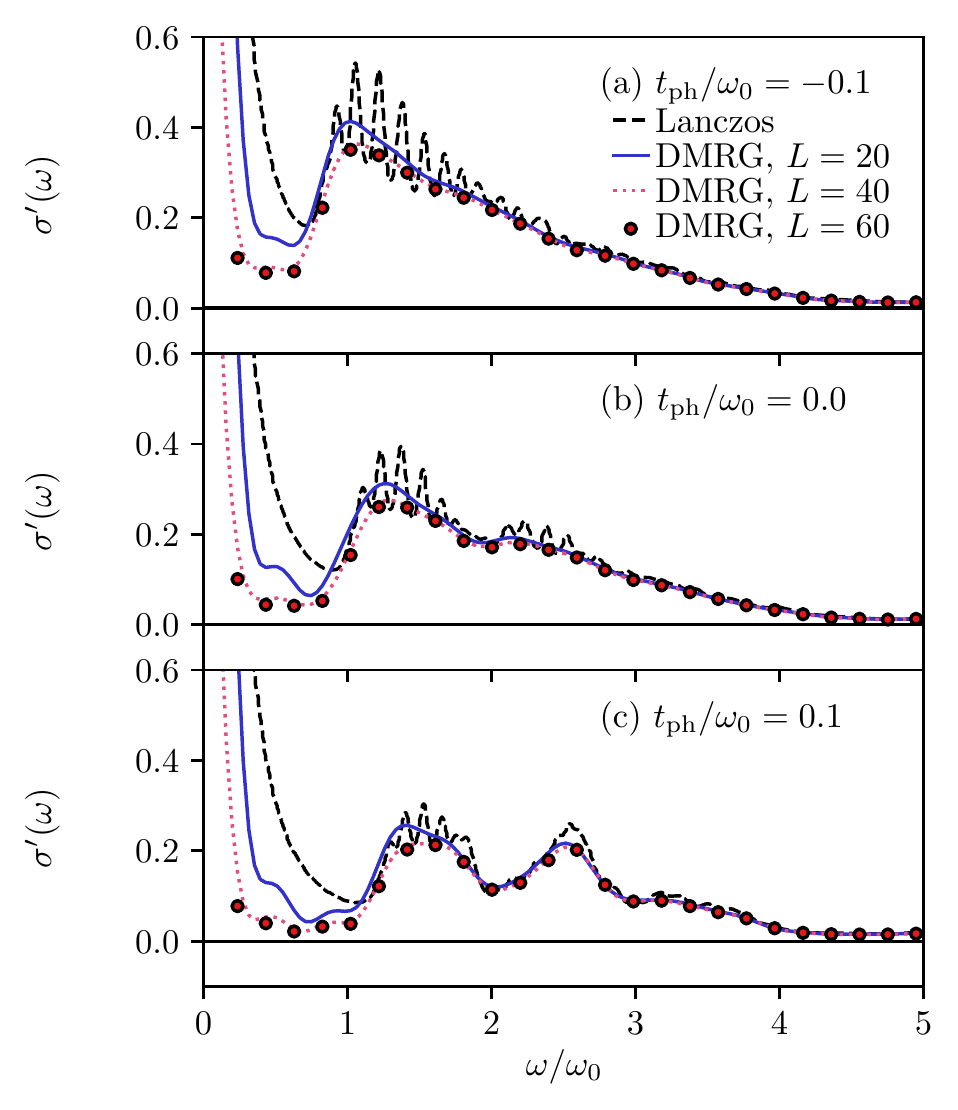}
 \caption{(a) Real part of the optical conductivity for the Holstein polaron in the ground state for $\lambda=1/2$, $M=20,\epsilon_{\rm  Bond}=\epsilon_{\rm LBO}=10^{-9}$, and with $t_{\rm{ph}}/\omega_0=-0.1$.  (b) Same as in (a) but with   and $t_{\rm{ph}}/\omega_0=0.0$.  (c) Same as in (a) but with $t_{\rm{ph}}/\omega_0=0.1$.  We show $L=60$, $40$, $20$,  and the time evolution is done up to $t_{\rm max }\omega_0=16$.  We further use a Lorenzian broadening with $\eta=0.1$. The black dashed line is the reference data computed with the Lanczos method.  The DMRG simulations, obtained with p2TDVP-LBO, are distributed onto 12,  8 and 4 processes for $L=60$, $40$, and $20$, respectively. }
 \label{fig:OPCGS10}
 \end{figure}

  In Fig.~\ref{fig:OPCGS10}, we show the real part of the optical conductivity in the Holstein-polaron model in the weak-coupling regime, $\lambda=1/2$, for different values of the phonon hopping amplitude $t_{\rm{ph}}$, calculated with the Lanczos method and p2TDVP-LBO for $L=20$, $40$ and $60$. The first important result is that the p2TDVP-LBO  method reproduces the incoherent part of  $\sigma^{\prime}(\omega)$ from the Lanczos method very accurately. Additionally,  we see that we can access system sizes large enough to obtain results approximately independent thereof. 

Figure~\ref{fig:OPCGS10}(b) shows the real part of the optical conductivity at finite $\omega$ and a flat phonon band. The incoherent part is dominated by two peaks starting at approximately $\omega / \omega_0=1$ and $2$. The appearance of a peak at $\omega / \omega_0=1$ is consistent with weak-coupling perturbation theory, see, e.g., Ref.~\cite{fratini_06},  and stems from when the one-phonon emission process.  The width of the peaks is also given by the possible changes in electron quasi-momenta $k_{\textrm{el}}$. This is because the emitted phonon can have any quasi-momentum due to the flat dispersion relation and the transition energies are thus dominated by the change in electron energy.  Here, the biggest difference between the Lanczos and the p2TDVP-LBO data is visible. The Lanczos data exhibit a peak structure occurring due to the possible electron quasi-momenta $k_{\rm el}$, which we cannot resolve with p2TDVP-LBO due to the finite resolution coming from the limited times available and the large number of $k_{\rm el}$.

When a small upward dispersion relation, $t_{\rm{ph}}/\omega_0=-0.1$, is introduced, see Fig.~\ref{fig:OPCGS10}(a),  the distinction between the different phonon emission peaks disappears and the weight of the incoherent spectra gets slightly shifted to lower frequencies. The latter can be explained by the fact that the emitted phonon has an additional energy contribution of $\approx 2t_{\textrm{ph}}$ due to the finite phonon bandwidth. As a result, the peak shifts to $\approx \omega_0+2t_{\textrm{ph}}$. The transfer to higher quasi-momentum states still remains rather unaffected since $\abs{t_0/t_{\rm{ph}}} \gg 1$.  Similar behavior of the second phonon peak leads to the monotonic decay of the incoherent spectra after the first maximum.

For $t_{\rm{ph}}/\omega_0=0.1$ in Fig.~\ref{fig:OPCGS10}(c), we see the opposite effect. The one- and two-phonon emission peaks become more easily separable and their maxima get shifted to larger frequencies. This is consistent with the fact that every transition energy introduces the term $2t_{\rm{ph}}\cos(k)$. For small $k$, the energy differences correspond to larger frequencies whereas the opposite happens for larger $k$ with $2t_{\rm{ph}} \cos(k)>0$. In total, this leads to a shift of the peaks to higher frequencies and a suppression of their width.
  
\begin{figure}[t]
 \includegraphics[width=0.99\columnwidth, height=10cm]{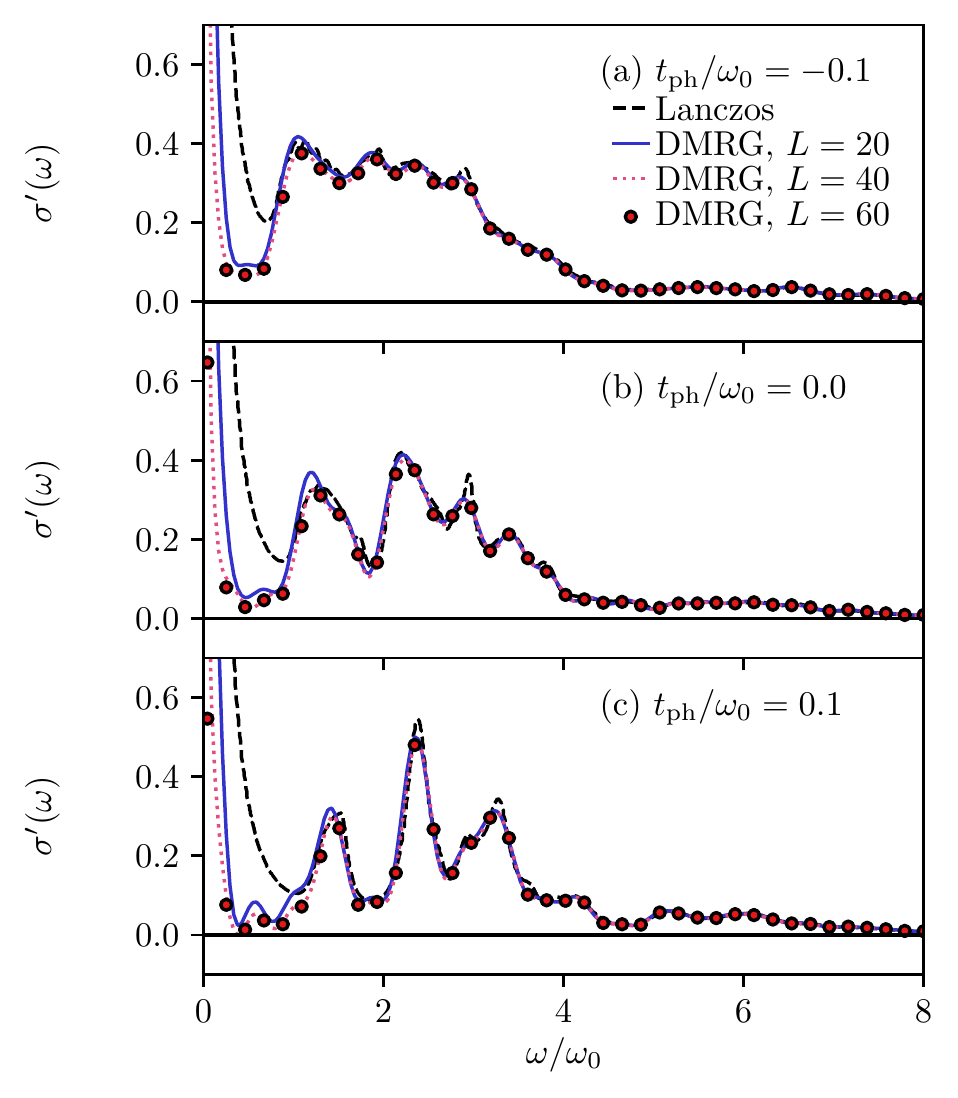}
 \caption{(a) Real part of the optical conductivity for the Holstein polaron in the ground state for $\lambda=1$, $M=20,\epsilon_{\rm  Bond}=\epsilon_{\rm LBO}=10^{-9}$, and with $t_{\rm{ph}}/\omega_0=-0.1$.  (b) Same as in (a) but with   and $t_{\rm{ph}}/\omega_0=0.0$.  (c) Same as in (a) but with $t_{\rm{ph}}/\omega_0=0.1$.  We show $L=60$, $40$, $20$,  and the time evolution is done up to $t_{\rm max }\omega_0=15$.  We further use a Lorenzian broadening with $\eta=0.08$. The black dashed line is the reference data computed with the Lanczos method.  The DMRG simulations, calculated using p2TDVP-LBO, are distributed onto 12,  8 and 4 processes for $L=60$, $40$, and $20$, respectively. }
 \label{fig:OPCGS14}
 \end{figure}

  We now turn to the intermediate coupling regime with $\lambda=1$. The ground-state results are presented in Fig.~\ref{fig:OPCGS14}. We show $t_{\rm{ph}}/\omega_0=-0.1$ in Fig.~\ref{fig:OPCGS14}(a), $t_{\rm{ph}}/\omega_0=0$ in Fig.~\ref{fig:OPCGS14}(b), and $t_{\rm{ph}}/\omega_0=0.1$ in Fig.~\ref{fig:OPCGS14}(c).  Again, essentially the same incoherent structure is produced by both the Lanczos and p2TDVP-LBO method with small differences in some peak heights. In this regime, a small but finite phonon bandwidth has a significant impact on the optical conductivity.  In general, the spectra broaden, signaling that the two and three phonon-emission processes play a more important role when the coupling is increased. Note that due to the small values for the dispersion, we do not observe the multi-phonon structure at $\omega<\omega_0$ for $t_{\rm{ph}}>0$, as reported in Ref.~\cite{bonca_21}. Still,  the same physical effects as in the weak-coupling case can be observed, although significantly enhanced. For example, the one- and two-phonon peak for $t_{\rm{ph}}/\omega_0=0.1$ in Fig.~\ref{fig:OPCGS14}(c) are almost completely distinguishable. In conclusion, we see that the p2TDVP-LBO method can reproduce all the features from the Lanczos method very well in the incoherent part of the spectrum for both $\lambda=1$ and $\lambda=1/2$. Note that we observe some oscillations for low frequencies which we attribute to artifacts of the Fourier transformation. This was verified by comparing the DMRG data to only the regular part of the Lanczos data.

   \label{subsec:opcGS}
\subsection{Ground state in the strong-coupling regime}
  \label{subsec:opcGSSC}
\begin{figure}[t]
\includegraphics[width=0.99\columnwidth, height=10cm]{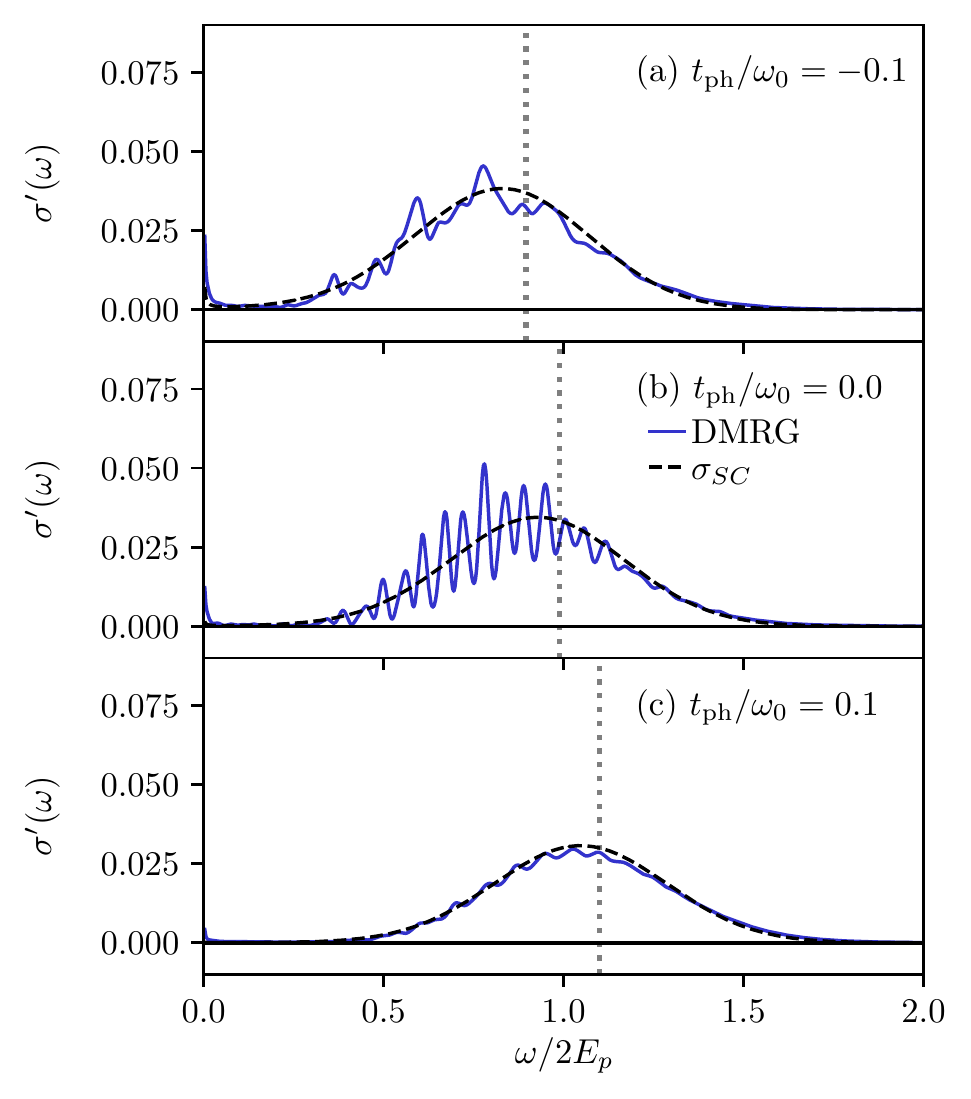}
\caption{(a) Real part of the optical conductivity for the Holstein polaron in the ground state for $\lambda=4.5$.  We further set $M=35$, $\epsilon_{\rm  Bond}=\epsilon_{\rm LBO}=10^{-7}$, $L=20$ and $t_{\rm{ph}}/\omega_0=-0.1$.  (b) Same as in (a) but with  $t_{\rm{ph}}/\omega_0=0$. (c) Same as in (a) but with $t_{\rm{ph}}/\omega_0=0.1$.  The blue solid line is the DMRG data and the dashed solid line is the analytical formula from Eq.~\eqref{eq:def_analyt_formSC}.  The vertical gray dashed lines show the respective energy differences between the Born-Oppenheimer surfaces at $\bar{q}_{\textrm{min},-}$. The time-evolution is done up to $t_{\rm max }\omega_0=11$ and  we use a Lorenzian broadening with $\eta=0.1$.  The simulation is distributed onto four processes for the DMRG calculations using p2TDVP-LBO. }
\label{fig:OPCGS30}
\end{figure}

   We now turn to the ground-state optical conductivity in the strong-coupling regime with $\gamma/\omega_0=3$.  As can be seen from Eq.~\eqref{eq:def_analyt_formSC} and by analyzing the Born-Oppenheimer surfaces in Fig.~\ref{fig:BOsurf_comb}, the expectation is a shifted Gaussian around twice the polaron binding energy $E_P$ for a flat phonon dispersion relation.  With a finite phonon bandwidth, one expects the Gaussian to be shifted to either larger or smaller frequencies depending on the sign of $t_{\rm{ph}}$. This central peak can be understood in terms of the Franck-Condon transition in the extreme adiabatic limit.  There, the motion of the electron is so fast compared to that of the phonons that the electron is excited without changing the phonon configuration with an energy difference of $2E_P$.  This process is sketched in Fig.~\ref{fig:pic}. In Fig.~\ref{fig:OPCGS30},  the results of the p2TDVP-LBO ground-state calculation for different phonon dispersion is plotted.  We show the expression from Eq.~\eqref{eq:def_analyt_formSC} and the p2TDVP-LBO data for $t_{\rm{ph}}/\omega_0=-0.1$ in Fig.~\ref{fig:OPCGS30}(a), $t_{\rm{ph}}/\omega_0=0$ in Fig.~\ref{fig:OPCGS30}(b), and $t_{\rm{ph}}/\omega_0=0.1$ in Fig.~\ref{fig:OPCGS30}(c).  Our numerical results are very well approximated by the analytic formula.
 Comparing Figs.~\ref{fig:OPCGS30}(a), (b), and (c), it becomes clear that even a small but finite phonon bandwidth significantly alters the position of the maximum of the spectrum. A downwards phonon dispersion relation shifts the absorption spectrum to higher frequencies and an upwards phonon dispersion relation shifts it to lower frequencies. This is expected from inspecting the Born-Oppenheimer surfaces in Sec~\ref{sec:BOopc}.  The vertical gray dashed lines in Fig.~\ref{fig:OPCGS30} indicate the energy differences between the two Born-Oppenheimer surfaces at the minimum of the lowest surface.
 
We conclude that our method also works well for a large electron-phonon coupling in the ground state, but we note that many local modes are needed in the local basis optimization. For example, the calculations in Figs.~\ref{fig:OPCGS10}, \ref{fig:OPCGS14}, and \ref{fig:OPCGS30}, required roughly 10, 15, and 30 local optimal basis states in both steps one and two of the algorithm. We also emphasize that, despite being calculated far from the adiabatic limit ($t_0/ \omega_0=1$), our results agree well with the analytic predictions. Note that in Fig.~\ref{fig:OPCGS30}, there is still a resilient M dependence in the amplitude of
some peaks, most notably so in Fig.~\ref{fig:OPCGS30}(b). This is further discussed in Appendix~\ref{sec:app2}.
\section{Finite-temperature results}
\label{sec:FTresults}

   \subsection{Finite-temperature results in the weak and intermediate coupling regimes}
  \label{subsec:opcFTWIC}
\begin{figure}[t]
\includegraphics[width=0.99\columnwidth, height=10cm]{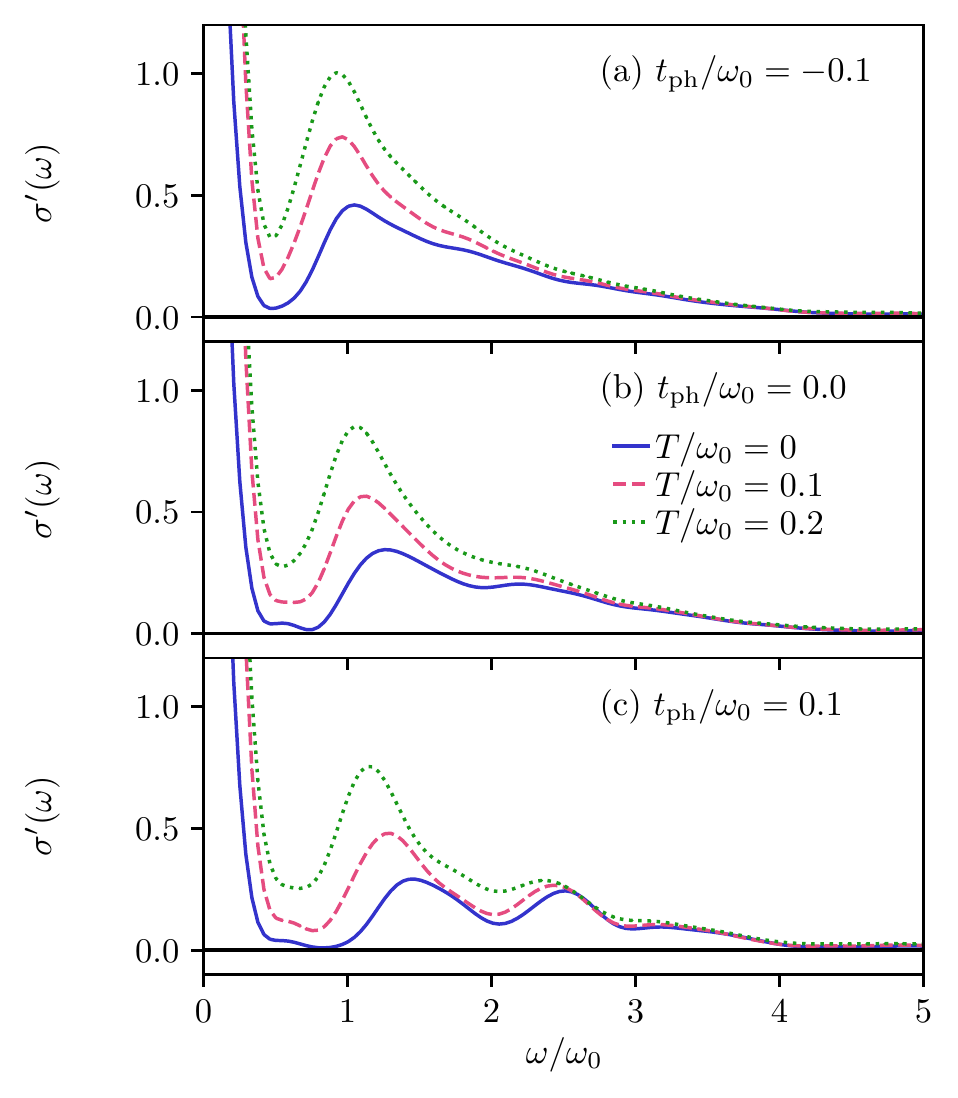}
\caption{(a) Real part of the optical conductivity at finite temperatures for $\lambda=1/2$ and with $t_{\rm{ph}}/\omega_0=-0.1$. (b) Same as in (a) but with $t_{\rm{ph}}/\omega_0=0$.  (c) Same as in (a) but with $t_{\rm{ph}}/\omega_0=0.1$. We further set $M=20,L=20$, and $\epsilon_{\rm  Bond}=\epsilon_{\rm LBO}=10^{-9}$.   The time evolution is done up to $t_{\rm max }\omega_0=15$ and we use a Gaussian broadening with $\eta=0.1/(4  \pi)$.  The simulations are carried out with p2TDVP-LBO and distributed onto 4 processes for the ground-state calculations and onto eight processes at finite temperature.}
\label{fig:OPCFT10}
\end{figure}

We proceed by investigating how the real part of the optical conductivity changes when going from zero to finite temperatures. Figure~\ref{fig:OPCFT10} shows $\sigma^{\prime}(\omega)$ for the Holstein polaron for $\lambda=1/2$ at different temperatures and with different values of the phonon hopping amplitude $t_{\rm{ph}}$. One can observe a similar influence of the finite temperature for the upwards dispersion relation [$t_{\rm{ph}}/\omega_0=-0.1$ in Fig.~\ref{fig:OPCFT10}(a)], a flat dispersion relation [$t_{\rm{ph}}/\omega_0=0$ in Fig.~\ref{fig:OPCFT10}(b)], and downwards dispersion relation [$t_{\rm{ph}}/\omega_0=0.1$ in Fig.~\ref{fig:OPCFT10}(c)]. Namely, there is a clear increase in spectral weight at low frequencies. This is consistent with other calculations in the weak-coupling regime~\cite{fratini_06} (although there in the adiabatic regime). This is due to the contribution of states previously thermally suppressed with a smaller energy difference than $\omega_0$. We further see an increase in the one-phonon emission-peak amplitude. This is due to the enhanced population of higher quasi-momentum states at a higher temperature, which now all contribute through one-phonon emission processes. The maximum one-phonon peak also shifts to lower frequencies, and the effect seems more apparent for $t_{\rm{ph}}/\omega_0=0.1$ than for $t_{\rm{ph}}/\omega_0=-0.1$. We attribute this to the fact that the sign of the curvature in the polaron and phonon bands differ in the first case and and is the same in the latter.
  
\begin{figure}[t]
\includegraphics[width=0.99\columnwidth, height=10cm]{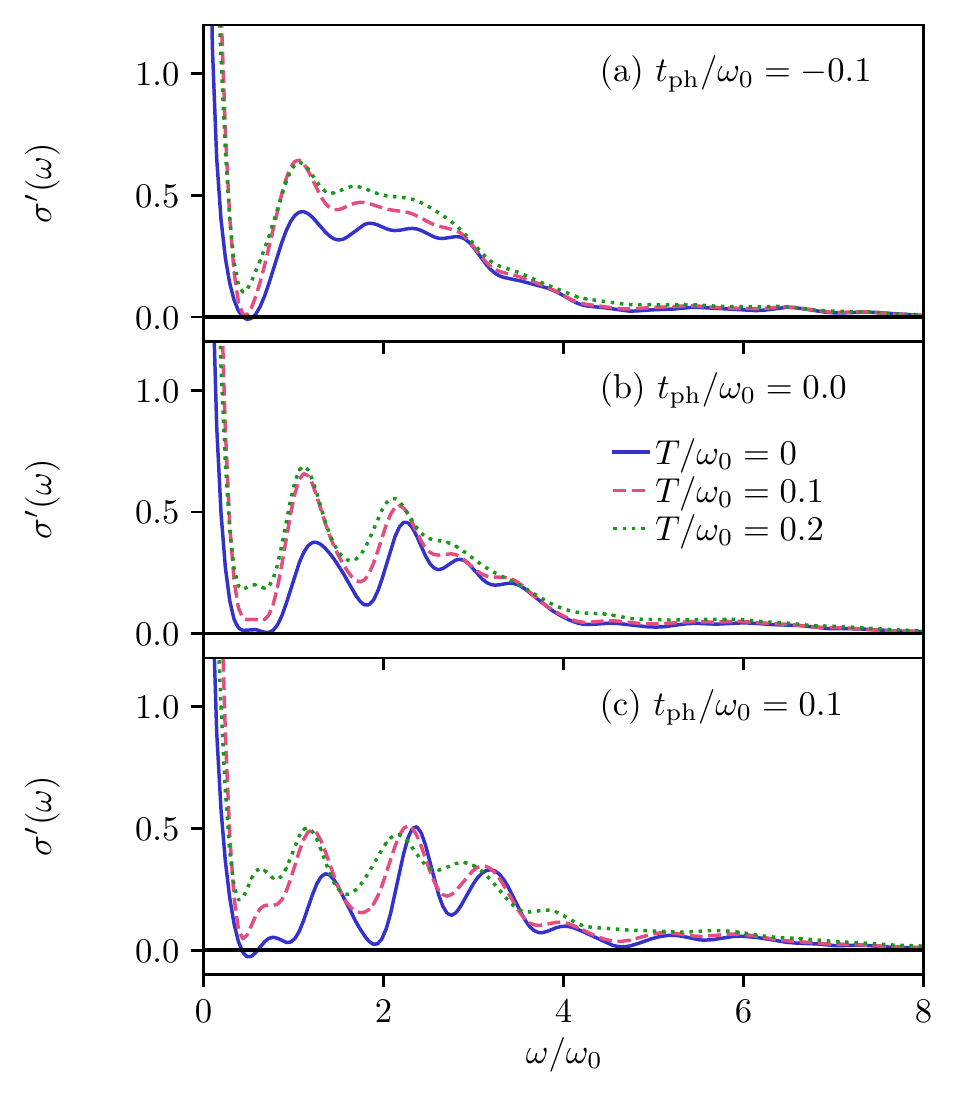}
\caption{(a) Real part of the optical conductivity at finite temperatures for $\lambda=1$ and with $t_{\rm{ph}}/\omega_0=-0.1$. (b) Same as in (a) but with $t_{\rm{ph}}/\omega_0=0$.  (c) Same as in (a) but with $t_{\rm{ph}}/\omega_0=0.1$. We further set $M=20,L=20$ and $\epsilon_{\rm  Bond}=\epsilon_{\rm LBO}=10^{-9}$ for $T/\omega_0=0$ and $\epsilon_{\rm  Bond}=5\cdot 10^{-9},\epsilon_{\rm LBO}=10^{-9}$  for  $T/\omega_0=0.1,0.2$. The time evolution is done up to $t_{\rm max }\omega_0=13$ and we use a Gaussian broadening with $\eta=0.1/(4  \pi)$.  The simulations are carried out with p2TDVP-LBO and distributed onto 4 processes for the ground-state calculations and onto  8 processes at finite temperature.}
\label{fig:OPCFT14}
\end{figure}

In the intermediate regime, $\lambda=1$ in Fig.~\ref{fig:OPCFT14}, there is a different influence of the finite temperature on the spectrum.  Whereas higher $T/\omega_0$ almost washes out the separate peak structure for $t_{\rm{ph}}/\omega_0=0$, $0.1$ and $\lambda=1/2$, the peaks remain fairly well separated for all temperatures studied here when $\lambda=1$.  This is valid for $t_{\rm{ph}}/\omega_0=0$, and $t_{\rm{ph}}/\omega_0=0.1$ in Fig.~\ref{fig:OPCFT14}(b) and (c), respectively.  Furthermore,  when $t_{\rm{ph}}/\omega_0=-0.1$,  the spectrum remains largely unaffected by temperature other than enhanced weight at lower frequencies. By decreasing $\eta$, we can observe small temperature dependent resonances at low frequencies for $\lambda=1/2$, $1$ and $t_{\textrm{ph}}/\omega_0=0$, $0.1$ [this can still be seen in Fig.~\ref{fig:OPCFT14}(c)], but from our available data, we cannot conclusively determine if they are physical or an artifact of the Fourier transformation. In this case, a higher frequency resolution would be necessary.
\subsection{Finite-temperature results in the strong-coupling regime}
  \label{subsec:opcFTSC}
\begin{figure}[t]
\includegraphics[width=0.99\columnwidth, height=10cm]{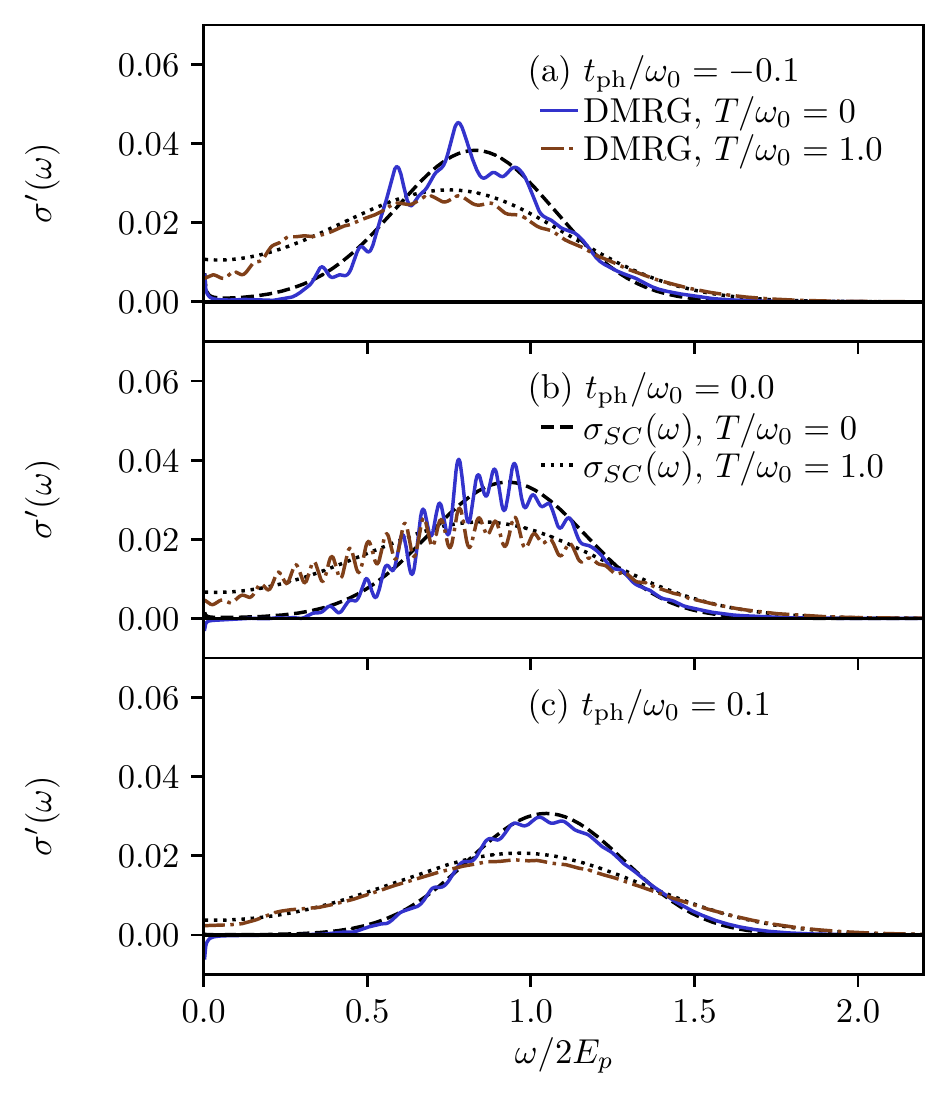}
\caption{(a) Real part of the optical conductivity at different temperatures for $\lambda=4.5$ and with $t_{\rm{ph}}/\omega_0=-0.1$. (b) Same as in (a) but with $t_{\rm{ph}}/\omega_0=0$. (c) Same as in (a) but with $t_{\rm{ph}}/\omega_0=0.1$.  We further set $M=35 $,  $L=20$ and $\epsilon_{\rm  Bond}=\epsilon_{\rm LBO}=10^{-7}$ for the ground-state calculations.  For the real-time evolution for $T/\omega_0=1.0$, we use single-site TDVP with $\epsilon_{\textrm{bond},J}=10^{-8}$ for $t_{\rm{ph}}/\omega_0=-0.1$, $0.1$ and  $\epsilon_{\textrm{bond},J}=10^{-9}$ for $t_{\rm{ph}}/\omega_0=0$ (see Sec.~\ref{sec:methods} for details). The time evolution is done up to $t_{\rm max }\omega_0=11$ and we use a Gaussian broadening with $\eta=0.5/(4\pi)$.  The black dashed and dotted lines are the analytical formula from Eq.~\eqref{eq:def_analyt_formSC}. The ground-state DMRG  results are distributed onto 4 processes and are calculated with p2TDVP-LBO. }
 \label{fig:OPCFT30}
 \end{figure}

We now analyze how the optical conductivity changes with temperature in the strong-coupling regime,  i.e., $\lambda=4.5$.  The results can be seen in Figs.~\ref{fig:OPCFT30}(a), \ref{fig:OPCFT30}(b), and \ref{fig:OPCFT30}(c) for the phonon dispersion $t_{\rm{ph}}/\omega_0=-0.1$, $0$, and $0.1$, respectively.  The analytic formula for the optical conductivity, see Eq.~\eqref{eq:def_analyt_formSC}, remains a good approximation in all cases in Fig.~\ref{fig:OPCFT30}, even as the temperature is increased to $T/\omega_0=1$.  Still, our data have less spectral weight at low frequencies. The reason is not obvious, but some deviations are to be expected 
since we are far away from the adiabatic limit. Further, we observe that the phonon dispersion relation affects the finite-temperature behavior.  The upwards dispersion relation, see Fig.~\ref{fig:OPCFT30}(a), leads to a larger contribution at low frequencies than the downward dispersion relation in Fig.~\ref{fig:OPCFT30}(c).  Looking at the $\omega\rightarrow 0$ values, it appears that the thermally activated hopping plays a lesser role when there is a downwards than an upwards phonon dispersion relation. This is consistent with the depth of the lower Born-Oppenheimer surfaces increasing, as can be seen in Fig.~\ref{fig:BOsurf_comb}.  Furthermore, since the low quasi-momenta phonons have higher energy for the downwards dispersion relation, the nondiagonal transition with a small change in electron quasi-momenta is suppressed compared to the upwards dispersion relation case.  Note that the resonance at $\omega=2t_0$ seen in Ref.~\cite{schubert_05}  for a small $\omega_0/t_0$ can be qualitatively reproduced by our method (not shown here). We further mention that small finite-size effects can be seen in some of the oscillation amplitudes in the data shown in Fig.~\ref{fig:OPCFT30}. This is illustrated in Appendix \ref{sec:app2}.
\section{Bipolaron results}
\label{sec:Bipolresults}
\begin{figure}[t]
\includegraphics[width=0.99\columnwidth]{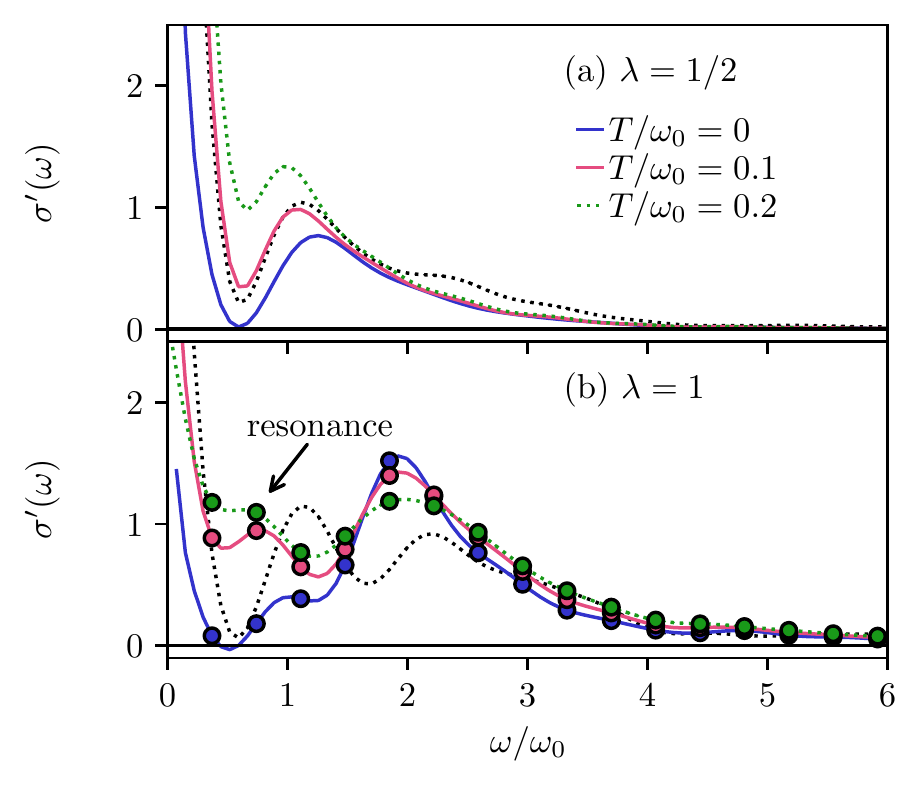}
\caption{(a) Optical conductivity for the Holstein bipolaron  for $\lambda=1/2$, $L=20$, $M=20$ and  $t_{\rm{ph}}/\omega_0=0$.  The black dotted line is the polaron curve with $\lambda=1/2$, $T=0.1$,  $\epsilon_{\textrm{bond}}=\epsilon_{\textrm{LBO}}=10^{-9}$ and scaled with a factor of two. (b) Same as (a) but with $\lambda=1$. We use $\epsilon_{\textrm{bond}}=5 \cdot 10^{-9}$, $\epsilon_{\textrm{LBO}}=10^{-9}$ for the ground-state data and $\epsilon_{\textrm{bond}}=5 \cdot 10^{-9}$, $\epsilon_{\textrm{LBO}}=10^{-8}$ for the finite-temperature data.  The symbols show the data points without zero padding. We time evolve to $t_{\rm max }\omega_0=8.5$ and use a Gaussian broadening with $\eta=0.2/(4\pi)$.  The black dotted line in (b) is the polaron curve with $\lambda=1$, $T=0.1$,  $\epsilon_{\textrm{bond}}=5\cdot 10^{-9}$,  $\epsilon_{\textrm{LBO}}=10^{-9}$ and scaled with a factor of two.  The simulations are done with p2TDVP-LBO and are distributed onto 4 processes for the ground-state calculations and onto (ten) eight processes for the (bi)polaron finite-temperature calculations. }
\label{fig:OPCFTbipolmix}
\end{figure}

We now proceed to study the real part of the optical conductivity of the Holstein bipolaron at finite temperature. In Fig.~\ref{fig:OPCFTbipolmix}, we plot $\sigma^{\prime}(\omega)$ for weak and intermediate electron-phonon coupling, $\lambda=1/2$, $1$, with $t_{\rm{ph}}/\omega_0=0$.  We observe substantial differences in the spectrum between both the two different electron-phonon coupling parameters, as well as between the polaron and bipolaron for $\lambda=1$. For $\lambda=1/2$, see Fig.~\ref{fig:OPCFTbipolmix}(a), the bipolaron spectrum is almost identical to the single-polaron spectrum weighted by a factor of two. When $\lambda=1$, we obtain quite a different picture. Whereas the $\omega/\omega_0=1$,  $2$ peaks have a similar amplitude for the polaron when $t_{\rm{ph}}/\omega_0=0$, see Fig.~\ref{fig:OPCFT14}(b), the bipolaron has a clear maximum around $\omega/\omega_0=2$. This indicates that the two-phonon emission process plays a much more dominant role than for the polaron. As $T/\omega_0$ is increased, a resonance appears that can be distinguished from the rest of the spectrum [indicated by the arrow in Fig.~\ref{fig:OPCFTbipolmix}(b)].  We believe that this stems from the enhanced importance of the one phonon emission peak, which shifts to lower frequencies, similar to what can be seen in Fig.~\ref{fig:OPCFTbipolmix}(a).  When the temperature is increased, the two electrons act more like separate polarons, and the one-phonon emission process becomes more important (for very high temperatures, the bipolaron will dissolve).  

We want to emphasize that we focus on the single-bipolaron limit to properly understand the behavior in a dilute system.  This is also motivated by the recent results on bipolaron high-temperature superconductivity reported in Ref.~\cite{zhang_sous_22}.  Going to a finite bipolaron density will add scattering,
whose effects are beyond the scope of our study. We also observe that when we increase the density to two spinless electrons,  the spectrum does practically not change compared to the single electron case (except for a re-scaling with a factor of two). This indicates that the change in the spectrum in Fig.~\ref{fig:OPCFTbipolmix}(b) can be attributed to the formation of a bound bipolaron and not to the change in density. In Fig.~\ref{fig:OPCFTbipolmix}(b), we also show the data points without zero padding to illustrate that the resonance is not an artifact of the Fourier transformation. While not shown here, we additionally confirmed that the ground-state data in Fig.~\ref{fig:OPCFTbipolmix} and data obtained with the Lanczos method agree, as we illustrate for the polaron in Sec.~\ref{sec:GSresults}.

\begin{figure}[t]
\includegraphics[width=0.99\columnwidth]{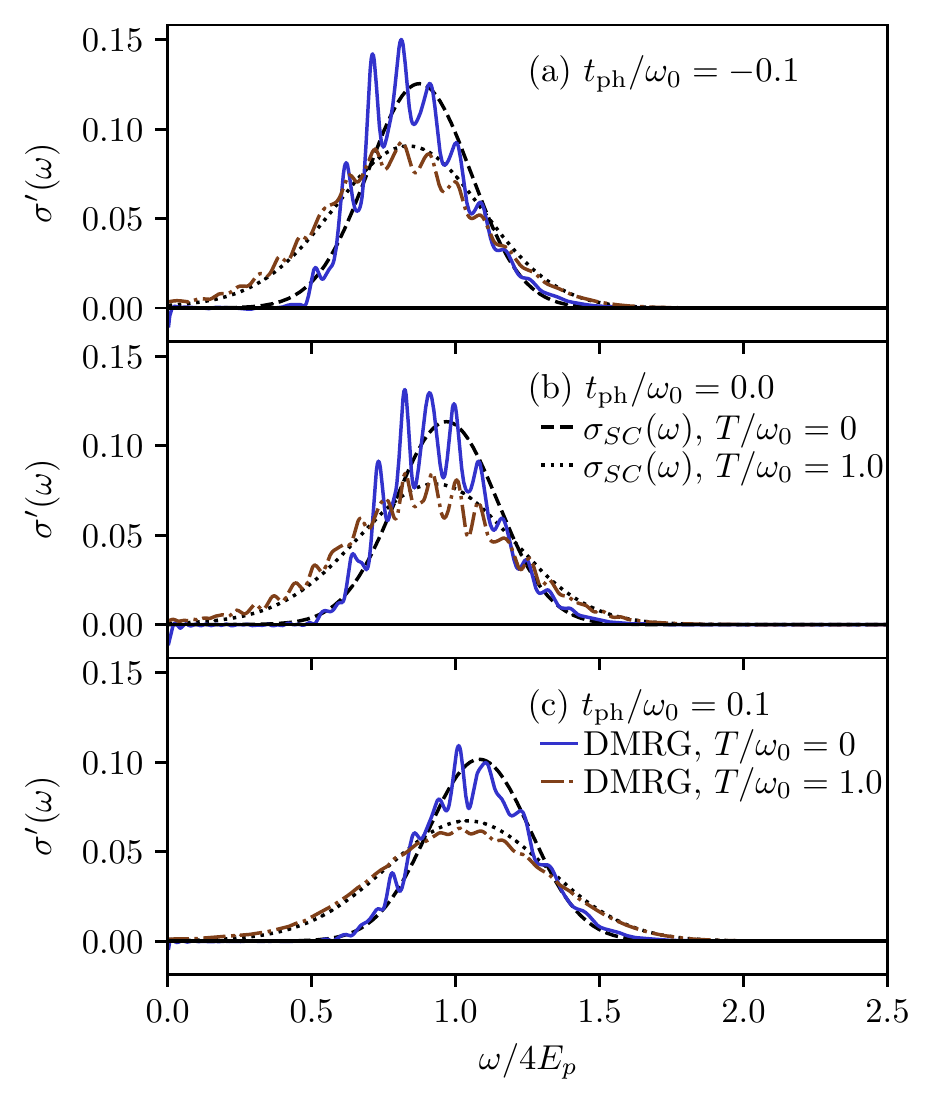}
\caption{(a) Optical conductivity for the Holstein bipolaron with $\lambda=4$ and $t_{\rm{ph}}/\omega_0=-0.1$. (b) Same as in (a) but with $t_{\rm{ph}}/\omega_0=0$. (c) Same as in (a) but with $t_{\rm{ph}}/\omega_0=0.1$. We further set $M=35,L=20$ and $\epsilon_{\rm  Bond}=\epsilon_{\rm LBO}=10^{-7}$ for the ground-state calculations.  For the time evolution for $T/\omega_0=1.0$, we use single-site TDVP with $\epsilon_{\textrm{bond},J}=10^{-7}$ for $t_{\rm{ph}}/\omega_0=-0.1$, $0.1$ and  $\epsilon_{\textrm{bond},J}=10^{-8}$ for $t_{\rm{ph}}/\omega_0=0$ (see Sec.~\ref{sec:methods} for details). The time-evolution is done up to $t_{\rm max }\omega_0=11$ and we use a Gaussian broadening with $\eta=0.2/(4\pi)$. The dashed and dotted black lines are the analytical formula from Eq.~\eqref{eq:def_analyt_formSC}. The ground-state simulations are done with p2TDVP-LBO and distributed onto 4 processes.}
\label{fig:OPCFTbipol20}
\end{figure}

In Fig.~\ref{fig:OPCFTbipol20}, we display the real part of the optical conductivity in the strong-coupling regime ($\lambda=4$). The first key observation is that the analytical formula in Eq.~\eqref{eq:def_analyt_formSC} is qualitatively consistent with the DMRG data. An asymmetric Gaussian with a dispersion-dependent maximum (around $4E_P$ for $t_{\rm{ph}}/\omega_0=0$) is seen in both cases. For $T/\omega_0=1$, the maximum is shifted to smaller frequencies and substantial spectral weight can be observed at small $\omega$.  This is similar to what occurred for the polaron,  see Fig.~\ref{fig:OPCFT30}, although a clear dispersion dependence on the low-frequency weight can not be identified for the parameters chosen here. Still, we conclude that Eq.~\eqref{eq:def_analyt_formSC} also describes the real part of the optical conductivity for the bound bipolaron quantitatively well. We note that we can still detect small finite-size effects in some of the oscillation amplitudes in Fig.~\ref{fig:OPCFTbipol20}. This is further discussed in Appendix \ref{sec:app2}.
\section{Conclusion}
\label{sec:conclusion}
In this work, we tackled the challenging task of accurately computing finite-temperature transport properties in polaron systems. We used single-site TDVP and two-site serial and parallel TDVP combined with local basis optimizations to calculate the real part of the optical conductivity for the Holstein polaron and bipolaron models with dispersive phonons at finite temperatures. We first explained how we incorporate LBO into two-site TDVP and reviewed the key ideas of the parallel implementation. The implementation of p2TDVP gives us the possibility to distribute the system onto different processes, thus allowing for an efficient time evolution while taking advantage of modern computer architectures. This, combined with LBO, allows us to compute time-dependent correlation functions for electron-phonon (bi)polaron systems with a variety of electron-phonon interaction strengths.

We first verified that the results obtained with p2TDVP are consistent with those calculated with the Lanczos method for the ground state of the Holstein polaron. We observed that even a small but finite phonon bandwidth has a significant impact on the optical conductivity for weak and intermediate electron-phonon coupling, consistent with previously reported results~\cite{bonca_21}. Most importantly, our method can accurately reproduce the Lanczos method data. For strong electron-phonon coupling, the known result of an asymmetric Gaussian is reproduced for $t_{\rm{ph}}/\omega_0=0$. However, a finite bandwidth shifts the center of the Gaussian to higher or lower frequencies depending on the sign of the phonon hopping amplitude $t_{\rm{ph}}$. This can be understood by inspecting the Born-Oppenheimer surfaces of the Holstein dimer as we saw in Sec.~\ref{sec:BOopc}.

Going to finite temperatures significantly alters the optical conductivity. For weak electron-phonon coupling, an increase of spectral weight around $\omega/\omega_0=1$ appears. Furthermore, the previously clearly distinguishable peaks almost completely merge at $t_{\rm{ph}}/\omega_0=0.1$.

For intermediate electron-phonon coupling, the separated peaks remain distinguishable for $t_{\rm{ph}}/\omega_0=0,0.1$, whereas the finite temperatures analyzed here do not alter the spectrum much for $t_{\rm{ph}}/\omega_0=-0.1$. In the large coupling case,  our data verify the validity for Eq.~\eqref{eq:def_analyt_formSC} in the non-adiabatic regime.  For $T/\omega_0=1$, the maximum of the Gaussian-shaped curve, compared to the $T/\omega_0=0$ case, shifts to smaller frequencies, and significant spectral weight is seen at lower frequencies. This is more prominent for an upwards than for a downwards phonon dispersion relation.

Lastly, we presented results for the bound bipolaron. There, we compared weak and intermediate electron-phonon coupling data for dispersionless phonons. While the real part of the optical conductivity for the first parameter set resembles that of the single polaron, a quite different behavior is seen when $\lambda=1$. There, the maximum of the spectrum is shifted to $\omega/\omega_0=2$, signaling the importance of two-phonon emission processes. These parameters also give rise to a large distinguishable peak at small frequencies. Our strong-coupling bipolaron data reproduce the analytic predictions drawn from the analysis of the Born-Oppenheimer surfaces very well. We reported a dispersion-dependent shift of the maximum of the spectrum and a significant decrease in the amplitude for $T/\omega_0=1$, in comparison to the zero-$T$ results. Furthermore, there is enhanced spectral weight for low frequencies.

These results characterize how the main features of small and large polarons in a Holstein-type system change under phonon dispersion and at finite temperatures. Understanding the basic mechanisms in such theoretical Hamiltonians constitutes an essential step in detecting such features, or lack thereof, in experimental data.

There are many interesting extensions to this work. For one, our numerical method can suffer from low-frequency resolution due to the limitations in reachable times. This could lead to small distinct features of the optical conductivity not being detected. Furthermore, despite the benefits of using LBO, there will always be a limit set by phonon truncation. For this reason, one goal of this work is to provide reliable data in the regimes where DMRG is the most powerful (intermediate electron-phonon interaction) which is out of reach of perturbative methods, see, e.g., Ref.~\cite{langfirsov}. These can then be used to reliably benchmark results obtained with other methods, see, e.g., Refs.~\cite{ehrenfest_27,tully_90,Shalashilin_09,Shalashilin_10,li13,mitric_21,Robinson_22_1,Alonso_2021,jancovich_22,brink_22}, which might have their strong suit in complementary parameter regimes.

Additionally, including electron-electron interactions in the Born-Oppenheimer formalism can lead to new insights into the transition from bound to unbound bipolarons~\cite{Fehske_94,Ihle_94,wellein96,bonca00,
Bonca_01,golez12a,werner_eckstein_15,marsiglio_22}. As this can involve additional numerical complexity, it can also serve as an interesting future application for the methods presented here.

 Furthermore, one could try to compare the computations with other finite-temperature matrix-product state methods such as minimally entangled typical thermal state algorithms (METTS), see, e.g., Refs. ~\cite{white2009,stoudenmire2010,binder_15}.

Lastly, an interesting application for our method would be to calculate properties for more complicated Hamiltonians, such as those inspired by manganite physics, see, e.g., Ref.~\cite{hotta_04} for a theoretical review.

\section*{Acknowledgment} \label{sec:ack}
We acknowledge useful discussions with P. Bl\"{o}chl,   C. Jooss, J. L\"{o}tfering, S. Manmana, C. Meyer, S. Paeckel, and M. ten Brink.
D.J. and F.H.-M. were  funded by the Deutsche Forschungsgemeinschaft (DFG, German Research Foundation) – 217133147 via SFB 1073 (Project B09). J.B.   acknowledges the support by the program P1-0044 of the
Slovenian Research Agency, support from the Centre for Integrated Nanotechnologies, a U.S. Department of Energy, Office of Basic Energy Sciences user facility, and funding from the Stewart Blusson Quantum Matter Institute. 
\\
\appendix
\section{Bond and local basis dimensions} \label{sec:app1}
    \begin{figure}[t]
\includegraphics[width=0.99\columnwidth]{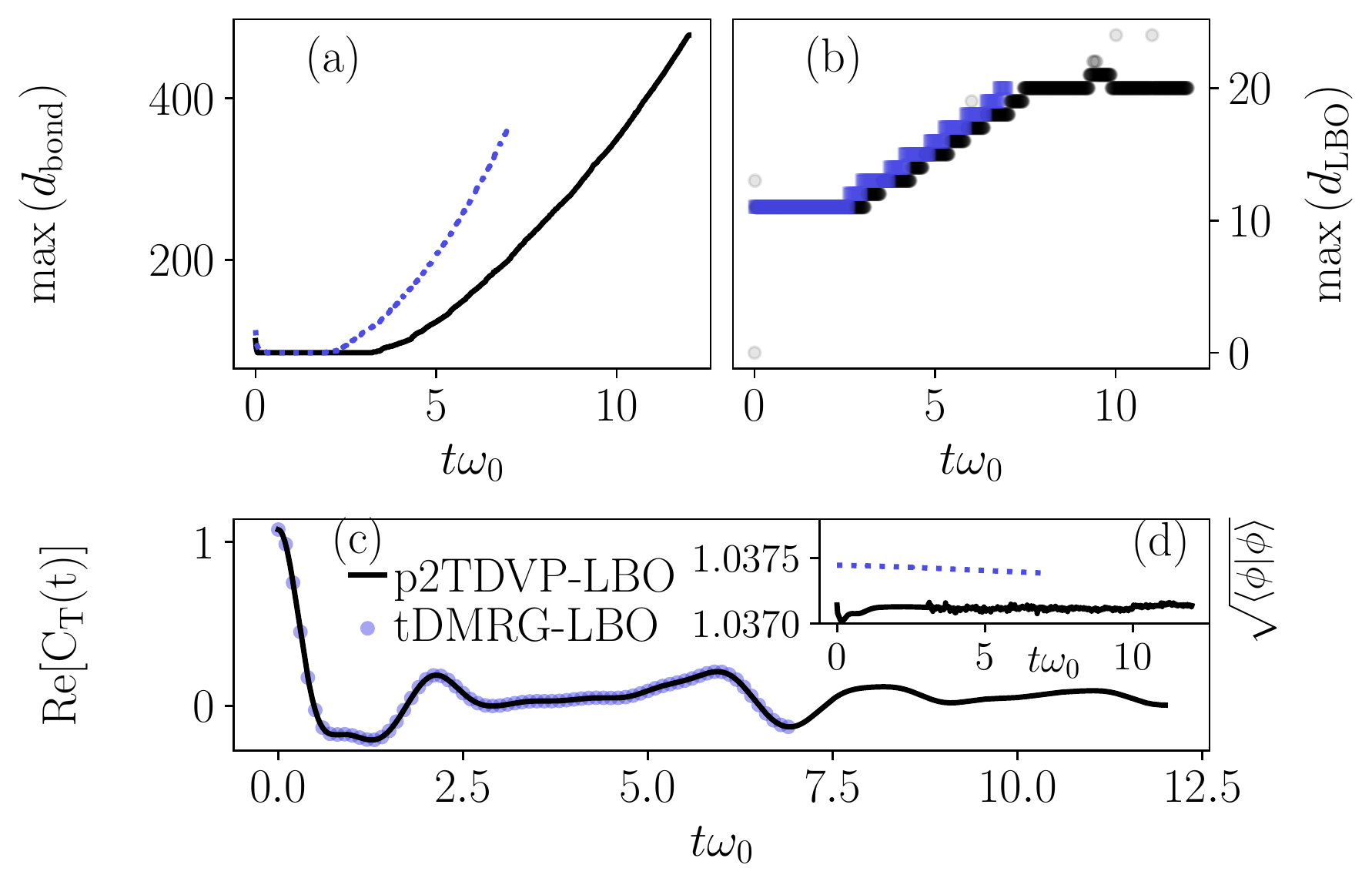}
\caption{(a) Maximum bond dimension of the matrix-product state at finite temperature after applying the current operator during the real-time evolution. We use with $\gamma / \omega_0=\sqrt{2}$, $t_{\rm{ph}}/\omega_0=0.1$, $M=20$, $L=20$, and $\epsilon_{\rm  Bond}=\epsilon_{\rm LBO}=10^{-9}$. The time-evolution methods are tDMRG-LBO and p2TDVP-LBO, see Appendix~\ref{sec:app1} for details. (b) Maximum local optimal basis dimension after the SVD in the time evolution algorithms (step two for p2TDVP-LBO and see Ref.~\cite{jansen20} for details for the tDMRG-LBO algorithm). (c) Real part of the current-current correlation function. (d) Norm of the initial state.  (b), (c) and (d) have the same parameters as (a).}
\label{fig:methcompBd}
\end{figure}
    \begin{figure}[t]
\includegraphics[width=0.99\columnwidth]{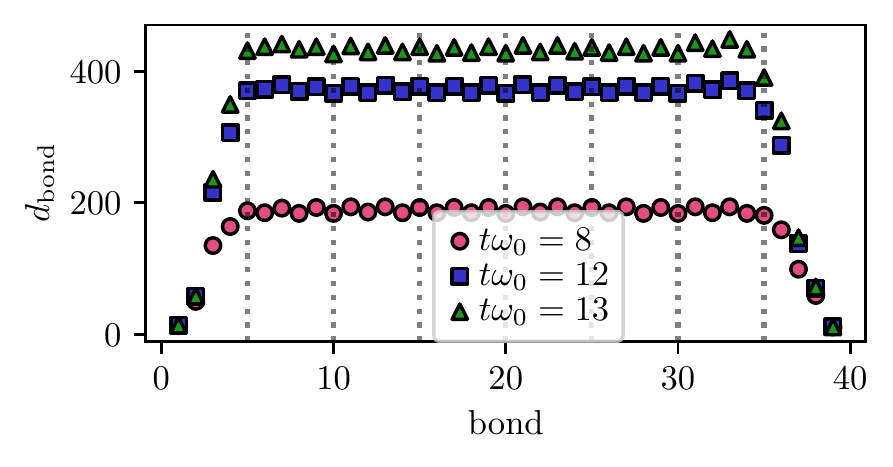}
\caption{Bond dimension between different sites at different times for the matrix-product state $\ket{\phi_T}$ for the Holstein polaron during time-evolution with p2TDVP-LBO. The calculations are done at the temperature $T/\omega_0=0.2$ and with the parameters $\lambda=1$, $t_{\rm{ph}}/\omega_0=0.1$, $M=20$, $L=20$, $\epsilon_{\rm  bond}=5\cdot 10^{-9},$ and  $\epsilon_{\rm LBO}=10^{-9}$. The dashed vertical lines show the partitioning of the system onto eight processes. }
\label{fig:mpsBD}
\end{figure}
We here discuss the numerical details of the p2TDVP-LBO algorithm. In Fig.~\ref{fig:methcompBd}, we compare how the computation of the current-current correlation functions scales compared to the tDMRG-LBO algorithm applied in Ref.~\cite{jansen20}. Note that we use tDMRG-LBO to obtain the thermal state with imaginary-time evolution and to do the real-time evolution shown in Fig.~\ref{fig:methcompBd}. The agreement between the two methods seen in Fig.~\ref{fig:methcompBd}(c) serves as a check for our p2TDVP-LBO algorithm. As explained in Sec.~\ref{sec:methods}, the real-time evolution is carried out on the state $\ket{\phi_T}=\hat{J}\ket{\psi_T}$.  In Fig.~\ref{fig:methcompBd}(a), we show the maximum bond dimension of $\ket{\phi_T}$ as a function of time with $\epsilon_{\rm  Bond}=\epsilon_{\rm LBO}=10^{-9}$ for both time-evolution methods. The need for a smaller bond dimension for p2TDVP is apparent, consistent with previous benchmarks of MPS time-evolution algorithms~\cite{paeckel_2019}. Figure~\ref{fig:methcompBd}(b) displays the maximum LBO dimension used in the SVD after applying the time-evolution gate in the tDMRG-LBO algorithm and step two of the p2TDVP-LBO algorithm (see Sec.~\ref{sec:methods} for details). In both cases, using LBO is beneficial for the parameters chosen here. In Fig.~\ref{fig:methcompBd}(d), we plot the norm of the state $\ket{\phi_T}$. Note that $\sqrt{\bra{\phi_T}\ket{\phi_T}}\neq 1$ since we already applied $\hat{J}$. The initial norm deviates between the two methods since the initial states are obtained using the different time-evolution schemes. Neither norm changes significantly during the reachable times. The reason why the time evolution is carried out until different times is the difference in computational cost. Note that this analysis does not account for the advantage of being able to distribute the computations onto several processes in the p2TDVP-LBO algorithm.

We now demonstrate that the calculations of the correlation function at finite temperatures can be homogeneously distributed onto several processes. In Fig.~\ref{fig:mpsBD}, we show that the bond dimension is roughly equal on all the bonds at different times for the bulk of the system. There, we plot the bond dimension between the different sites at the times $t\omega_0=8,12,13$ for the Holstein polaron. The parameters are given in the caption of Fig.~\ref{fig:mpsBD}. Furthermore, the grey dashed lines show where the partitioning of the system is done. Therefore,  a need for dynamically adapting the boundaries as suggested for pDMRG in Ref.~\cite{Chen_2021} does not seem necessary for our applications.

Lastly, in our discussion of the numerical details of p2TDVP-LBO, we illustrate the difference between using a regular Krylov solver, a Krylov solver with LBO (Krylov-LBO), and RK4 with LBO (RK4-LBO) in step one of the algorithm (see Sec.~\ref{sec:methods} for details). In our test case, we apply the three versions of the algorithm to the real-time evolution of the bipolaron ground state after applying the current operator $\ket{\phi_0}=\hat{J}\ket{\psi_0}$. Note that the results, as in all LBO calculations, will be system and set-up dependent and a benchmark should be carried out in each case separately when feasible.  

\begin{figure}[t]
\includegraphics[width=0.99\columnwidth]{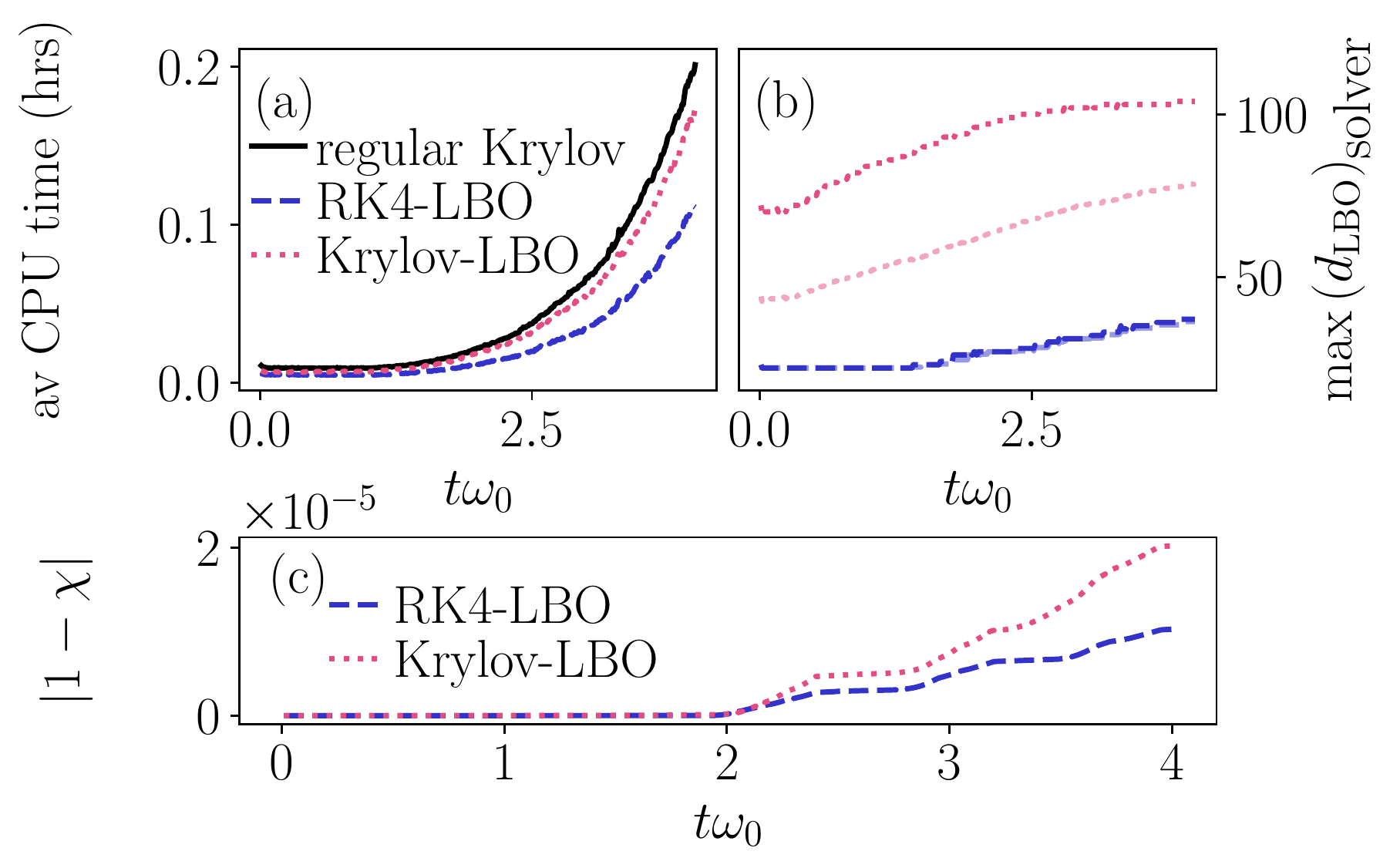}
\caption{Different local basis optimization schemes for two-site TDVP when calculating the real-time evolution of the bipolaron ground state after applying the current operator. The parameters are $L=13$, $M=25$, $\gamma/\omega_0=1$,  $t_{\textrm{ph}}/\omega_0=0.0$, and $\epsilon_{\rm  bond}=\epsilon_{\rm LBO}=10^{-9}$. (a) Average CPU time for the different implementations in step one of the two-site TDVP algorithm, see Sec.~\ref{sec:methods} for details. Black solid line: regular Krylov solver. Red dotted line: Krylov-LBO. Blue dashed line: RK4-LBO. (b) Maximum optimal basis dimension in the two-site TDVP with the Krylov-LBO (red dotted line) and with the RK4-LBO (blue dashed line) solver.  The light lines are calculated by averaging the LBO dimension in all the Krylov vectors or all the $k^j$ in the RK4 algorithm, and then taking the maximum of all the sites. (c) Absolute value of the overlap $\chi -1$ of the wave function with a LBO solver and with the regular Krylov solver.  Red dotted line: Krylov-LBO. Blue dashed line: RK4-LBO. All simulations use two cores.}
\label{fig:diffTDVP}
\end{figure}

Figure~\ref{fig:diffTDVP}(a) shows the average CPU time for each time step with the three methods. One first observes that the exponential increase in the bond dimension dominates the computation time in all cases. Still, using Krylov-LBO seems to be faster than using the regular Krylov solver. The RK4-LBO algorithm is significantly faster for the times reached here. We believe that this can be understood as follows: The Krylov vectors for a general state $\ket{\psi}$, are $\ket{\psi}_0=\ket{\psi}$, $\ket{\psi}_1=\hat H \ket{\psi}_0$, $\ket{\psi}_2=\hat H \ket{\psi}_1= \hat{H}^2 \ket{\psi}_0$, and so on. 
The LBO basis is found before $\hat H$ is applied. Through the polynomial application of $\hat H$, we expect a larger phonon Hilbert space to be explored, which requires a larger number of LBO states. However, for the RK4 algorithm, $\ket{k^1}=\hat H\ket{\psi}$, but $\ket*{k^2}=\hat H(\ket{\psi}+\alpha dt\ket*{k^1})=(\hat H\ket{\psi}+\alpha dt\hat H^2 \ket{\psi})$ etc, where $\alpha$ is a number. 
Now, the term including $\hat H^2$ is scaled with $dt$. We thus believe that for small $dt$, the optimal basis changes less drastically than for the Krylov vector $\ket{\psi}_2$.  This seems plausible when inspecting the LBO dimensions needed in the calculations. 

In Fig.~\ref{fig:diffTDVP}(b), we show the maximum LBO dimension needed in the different algorithms. The red dotted line is the maximum LBO dimension used in the Krylov-LBO steps whereas the light red dotted line is the average. The maximum LBO dimension in the Krylov algorithm quickly reaches the maximum, which we already attributed to the fact that one must apply the effective Hamiltonian several times. The average remains somewhat lower. The required LBO dimension is significantly smaller in RK4-LBO (the average and maximum almost perfectly overlap), making the algorithm more efficient. 

We also observe that the difference between RK4-LBO and Krylov-LBO is negligible for the time scales studied here.  Figure \ref{fig:diffTDVP}(c) shows the absolute value of $1-\chi$, where $\chi$ is the overlap between either the state using Krylov-LBO (red dotted line) or RK4-LBO (blue dashed line) with the state which uses a Krylov method without LBO. The difference is small for all times in our test case.
We further want to point out that the RK4 solver, in general, is not as stable as the Krylov solver. For example, it did not converge well for the imaginary-time evolution in our calculations (which is why we use the regular Krylov solver in that case). For that reason, we also compared our results computed with the RK4-LBO solver to those obtained with a Krylov solver without LBO. 

\section{Comparing the moments} \label{sec:app2}
\begin{figure}[t]
     \includegraphics[width=0.99\columnwidth]{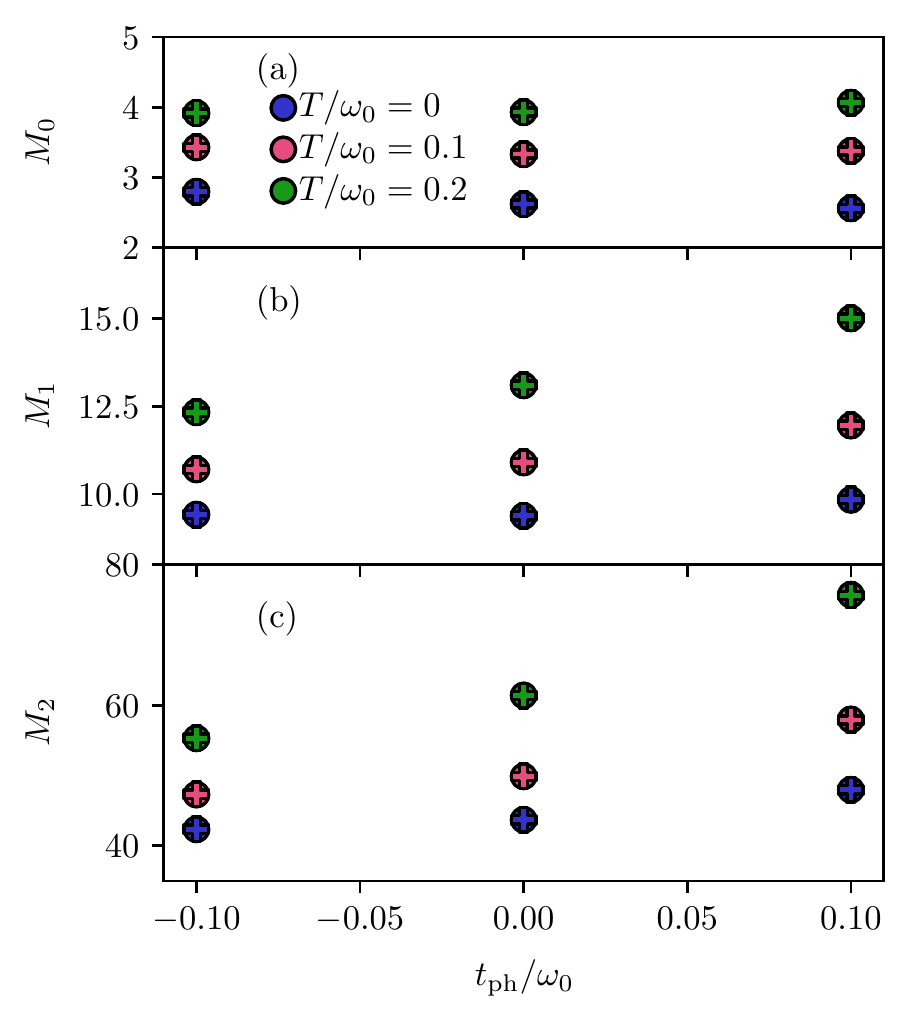}
\caption{Moments of the correlation function for the same parameters as in Fig.~\ref{fig:OPCFT14}.  (a) Zeroth moment, (b) first moment, (c) second moment.  The circles are calculated by integrating the Fourier transformed data obtained with time evolution and the plus signs from evaluating the commutators at finite temperatures. For details, see Appendix~\ref{sec:app2}. }
\label{fig:moms14}
\end{figure}

We now want to estimate the accuracy of the Fourier transform of the time-evolved current-current correlation function, see Eq.~\eqref{eq:def_corel0}, which we use to extract $\sigma^{\prime}(\omega)$. To do that, we compare the moments of the correlation function calculated in two ways: By calculating a thermal expectation value of the initial state and integrating the correlation function over $\omega$. Computing the moments to assess the accuracy of the calculations for the Holstein polaron has already been used for the spectral functions, see, e.g., Refs.~\cite{Kornilovitch_2002,goodvin06,bonca2019, jansen20}.
We define the $k$th moment to be 
\begin{equation}
 \label{eq:k_mom}
\tilde{M}_k=\int_{-\infty}	^{\infty}d \omega (\omega)^k C(\omega)_T \, .
\end{equation}
The zeroth moment is then:
\begin{equation}
 \label{eq:zero_mom}
\tilde{M}_0=\int_{-\infty}	^{\infty}d \omega C(\omega)_T=\pi \expval*{\hat J^2}_T \, .
\end{equation} 
The first moment becomes
\begin{equation}
 \label{eq:first_mom_1}
\tilde{M}_1=\int_{-\infty}	^{\infty}d \omega \omega C(\omega)_T=\pi  \expval*{[\hat{J},\hat{H}]\hat{J}}_T\, ,
\end{equation} 
where $[\hat{J},\hat{H}]=[\hat{J},\hat{H}_{\rm kin}]+[\hat{J},\hat{H}_{\rm e-ph}]$  with
\begin{equation}
 \label{eq:first_mom_2}
[\hat{J},\hat{H}_{\rm kin}]=-it_0^2 \sum\limits_{s} (2 {\hat{c}^{\dagger}_{1,s}\hat{c}_{1,s}} - 2 {\hat{c}^{\dagger}_{L,s}\hat{c}_{L,s}})
\end{equation}
and 
\begin{equation}
 \label{eq:first_mom_3}
 \begin{split}
[\hat{J},\hat{H}_{\rm e-ph}]&=\gamma t_0 i\bigg( \sum\limits_{j=2,s}^{L} (\hat{c}^{\dagger}_{j-1,s}\hat{c}_{j,s}+\hat{c}^{\dagger}_{j}\hat{c}_{j-1,s})(\hat{b}^{\dagger}_j+\hat{b}_j)  \\&
 - \sum\limits_{j=1,s}^{L-1} (\hat{c}^{\dagger}_{j,s}\hat{c}_{j+1,s}+ \hat{c}^{\dagger}_{j+1,s}\hat{c}_{j,s})(\hat{b}^{\dagger}_j+\hat{b}_j)\bigg)\, .
 \end{split}
\end{equation}
Lastly, the second moment becomes \begin{equation}
 \label{eq:sec_mom_1}
\tilde{M}_2=\int_{-\infty}	^{\infty}d \omega \omega^2 C(\omega)_T=\pi \expval*{[[\hat{J},\hat{H}],\hat{H}] \hat{J}}_T\, ,
\end{equation} 
with 
\begin{multline}
 \label{eq:sec_mom_2}
[[\hat{J},\hat{H}],\hat{H}]=it_0^3 2 \Biggl( \sum\limits_{s={\uparrow, \downarrow}}\Bigl(\hat{c}^{\dagger}_{1,s}\hat{c}_{2,s} -\hat{c}^{\dagger}_{2,s}\hat{c}_{1,s} 
\\ +\hat{c}^{\dagger}_{L-1,s}\hat{c}^{\dagger}_{L,s} -\hat{c}^{\dagger}_{L,s}\hat{c}_{L-1,s}\Bigr)\Biggr) 
\\ 
+\gamma t_0 \omega_0 i \Biggl(  \sum\limits_{j=2,s={\uparrow, \downarrow}}^{L} \Bigl(\hat{c}^{\dagger}_{j-1,s}\hat{c}_{j,s}\hat{b}_j+\hat{c}^{\dagger}_{j,s}\hat{c}_{j-1,s}\hat{b}_j\\
 -\hat{c}^{\dagger}_{j-1,s}\hat{c}_{j,s}\hat{b}^{\dagger}_j -\hat{c}^{\dagger}_{j,s}\hat{c}_{j-1,s}\hat{b}^{\dagger}_j \Bigr) 
 \\ + \sum\limits_{j=1,s={\uparrow, \downarrow}}^{L-1} \Bigl(-\hat{c}^{\dagger}_{j,s}\hat{c}_{j+1,s}\hat{b}_j- \hat{c}^{\dagger}_{j+1,s}\hat{c}_{j,s}\hat{b}_j\\
 +\hat{c}^{\dagger}_{j,s}\hat{c}_{j+1,s}\hat{b}^{\dagger}_j+ \hat{c}^{\dagger}_{j+1,s}\hat{c}_{j,s}\hat{b}^{\dagger}_j\Bigr) \Biggr) 
 \\ + it_0 \gamma t_{\textrm{ph}}  \Biggl( \sum\limits_{j=2,s={\uparrow, \downarrow}}^{L-1} \Bigl(\hat{c}^{\dagger}_{j-1,s}\hat{c}_{j}\hat{b}_{j+1}+\hat{c}^{\dagger}_{j,s}\hat{c}_{j-1,s}\hat{b}_{j+1}
\\ -\hat{c}^{\dagger}_{j-1,s}\hat{c}_{j,s}\hat{b}_{j+1}^{\dagger} -\hat{c}^{\dagger}_{j,s}\hat{c}_{j-1,s}\hat{b}_{j+1}^{\dagger} \Bigr) \Biggr)  \\
 -it_0 \gamma t_{\textrm{ph}}\Biggl( \sum\limits_{j=2,s={\uparrow, \downarrow}}^{L-1}\Bigl(\hat{c}^{\dagger}_{j,s}\hat{c}_{j+1,s}\hat{b}_{j}+ \hat{c}^{\dagger}_{j+1,s}\hat{c}_{j,s}\hat{b}_{j}
\\ -\hat{c}^{\dagger}_{j,s}\hat{c}_{j+1,s}\hat{b}_{j}^{\dagger}- \hat{c}^{\dagger}_{j+1,s}\hat{c}_{j,s}\hat{b}_{j}^{\dagger}\Bigr)\Biggr) \\
-i t_0 \gamma^2\Bigl(  \sum\limits_{j=1,s={\uparrow, \downarrow}}^{L-1} \Bigl( \hat{c}^{\dagger}_{j+1,s}\hat{c}_{j,s}(\hat{X}_j)^2- \hat{c}^{\dagger}_{j+1,s}\hat{c}_{j,s}\hat{X}_j\hat{X}_{j+1}
 \\ + \hat{c}^{\dagger}_{j,s}\hat{c}_{j+1,s}\hat{X}_j\hat{X}_{j+1}- \hat{c}^{\dagger}_{j,s}\hat{c}_{j+1,s}(\hat{X}_j)^2\Bigr)  \Biggr) 
\\
+i t_0 \gamma^2 \Biggl(  \sum\limits_{j=2,s={\uparrow, \downarrow}}^{L} \Bigl(\hat{c}^{\dagger}_{j-1,s}\hat{c}_{j,s}(\hat{X}_j)^2- \hat{c}^{\dagger}_{j-1,s}\hat{c}_{j,s}\hat{X}_j\hat{X}_{j-1}
\\ 
+  \hat{c}^{\dagger}_{j,s}\hat{c}_{j-1,s}\hat{X}_j\hat{X}_{j-1}- \hat{c}^{\dagger}_{j,s}\hat{c}_{j-1,s} (\hat{X}_j)^2 \Bigr) \Biggr) 
\\ -\gamma t_0^2 i \Biggl( \sum\limits_{j=1,s={\uparrow, \downarrow}}^{L-2} \Bigl( \hat{c}^{\dagger}_{j+2,s}\hat{c}_{j,s} -\hat{c}^{\dagger}_{j,s}\hat{c}_{j+2,s}
\Bigr) \hat{X}_j \\ 
 +\sum\limits_{j=1,s={\uparrow, \downarrow}}^{L-2} \Bigl( \hat{c}^{\dagger}_{j,s}\hat{c}_{j+2,s} -\hat{c}^{\dagger}_{j+2,s}\hat{c}_{j,s} \Bigr) \hat{X}_{j+1}\Biggr) 
 \\ -\gamma t_0^2 i \Biggl( \sum\limits_{j=2}^{L-1}\Bigl(\hat{c}^{\dagger}_{j-1,s}\hat{c}_{j+1,s}-\hat{c}^{\dagger}_{j+1,s}\hat{c}_{j-1,s} 
\Bigr) \hat X_j \\  
+\sum\limits_{j=2,s={\uparrow, \downarrow}}^{L-1}\Bigl( \hat{c}^{\dagger}_{j+1,s}\hat{c}_{j-1,s}-\hat{c}^{\dagger}_{j-1,s}\hat{c}_{j+1,s} \Bigr) \hat{X}_{j+1}\Biggr) \, ,
\end{multline}
with $\hat{X}_i=\hat{b}^{\dagger}_i+\hat{b}_i$.  Figure~\ref{fig:moms14} shows $\tilde{M}_1,\tilde{M}_2$ and $\tilde{M}_3$ for the same data as plotted in Fig.~\ref{fig:OPCFT14}. We display both the integrated correlation functions (integrated for $\omega/\omega_0 \in [-15,15] $) and the thermal expectation values.  The relative difference is at maximum of the order of $ \mathcal{O}(10^{-3})$.  For the data shown in Fig.~\ref{fig:OPCFT10}, the relative difference is also at maximum of the order of $ \mathcal{O}(10^{-3})$. The moments without zero padding are in some cases  just as good as for the data with zero padding.  For the DMRG data in Fig.~\ref{fig:OPCFT30}, we integrated for $\omega/ \omega_0  \in [-80,80] $ and obtain a relative difference at maximum of the order $ \mathcal{O}(10^{-3})$. For the bipolaron data in Fig.~\ref{fig:OPCFTbipolmix}, the difference in the moments are at most of the order $ \mathcal{O}(10^{-3})$ when integrated for $\omega/ \omega_0  \in [-15,15] $.  In Fig.~\ref{fig:OPCFTbipol20}, the relative difference is of order $ \mathcal{O}(10^{-3})$ when we integrate for $\omega/ \omega_0  \in [-50,50] $.

\begin{figure}[t]
     \includegraphics[width=0.99\columnwidth]{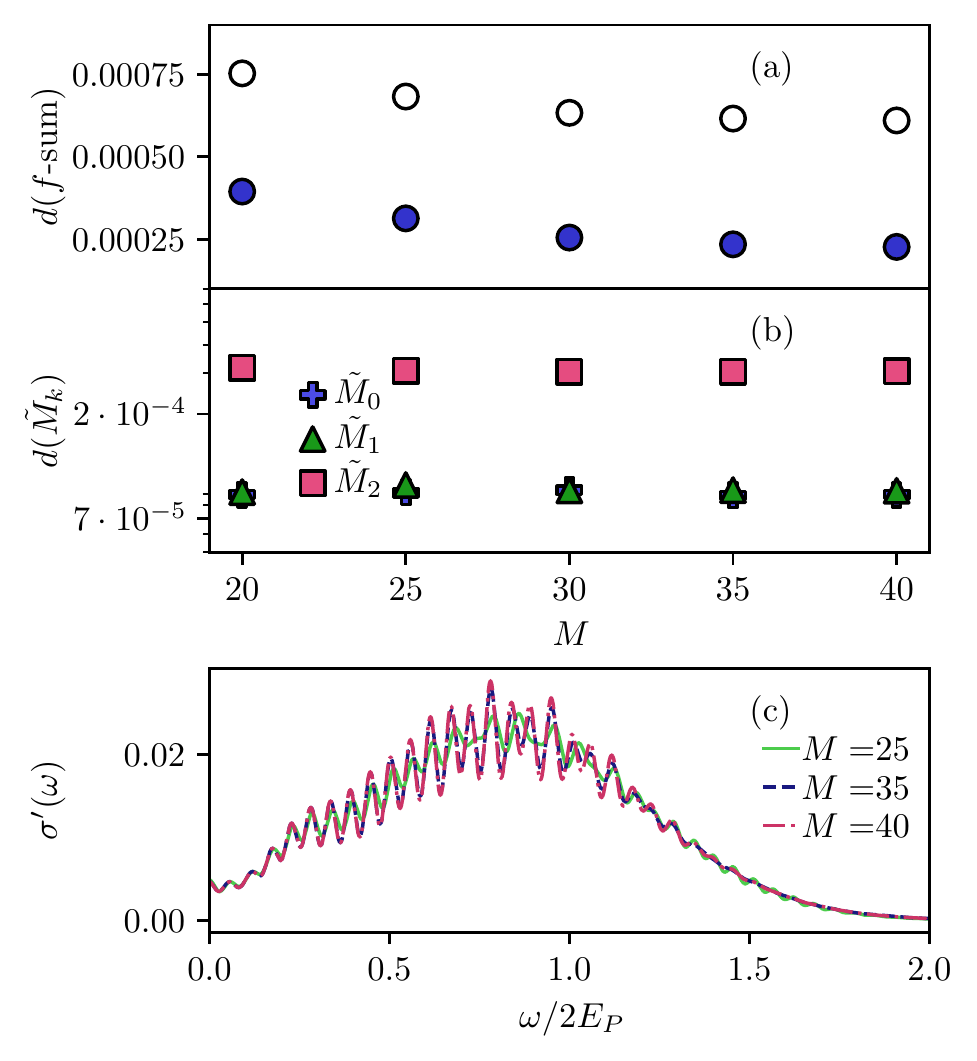}
\caption{(a) Relative difference in the $f$-sum rule, see Appendix~\ref{sec:app2} for details. (b) Relative difference in the moments, also see Appendix~\ref{sec:app2}.  (c) Real part of the optical conductivity.  All data are at $T/\omega_0=1$ and with different phonon cutoff $M$.  The rest of the parameters are the same as in Fig. ~\ref{fig:OPCFT30}.  The open circles in (a) show the data when no zero padding is used.}
\label{fig:moms_comp}
\end{figure}

To quantify the influence of the phonon cutoff $M$ for large coupling and large temperatures ($T/\omega_0=1$) we compare several quantities in Fig.~\ref{fig:moms_comp}. Figure~\ref{fig:moms_comp}(a) shows the relative difference in the $f$-sum rule between the initial state and the integrated real part of the optical conductivity: \begin{equation}
\label{eg:fsum}
d(\textrm{$f$-sum})=\frac{|\int_0^{\infty}d\omega \sigma^{\prime}(\omega)-\frac{-\pi\expval*{E_{\textrm{kin}}}_T}{2}|}{|\frac{-\pi\expval*{E_{\textrm{kin}}}_T}{2}|}\, .
\end{equation}
Note that all terms in Eq.~\eqref{eg:fsum} are evaluated with the same phonon truncation $M$ and that we integrate for $\omega/ \omega_0  \in [0,80] $.
We observe that the result improves as $M$ is increased. The empty symbols show the same data but without zero padding. As can be seen, the sparse number of integration points produces inaccurate results. We also mention that the type of methods used has difficulties correctly capturing the low-frequency physics due to the limitations in reachable times.  This can also lead to inaccurate results for the $f$-sum rule, in particular for systems with a significant Drude weight. 
In Fig.~\ref{fig:moms_comp}(b), we illustrate how the moments behave when $M$ is varied. There, we show the relative difference between the integral over the correlation function (integrated for $\omega/ \omega_0  \in [-80,80] $) and the thermal expectation value (i.e., right hand side of Eq.~\eqref{eq:zero_mom} for $\tilde{M}_0$) denoted as $\expval*{\hat{\tilde{M}}_k}_T$:
 \begin{equation}
\label{eg:dm}
d(\tilde{M}_k)=\frac{|\int_{-\infty}	^{\infty}d \omega (\omega)^k C(\omega)_T-\expval*{\hat{\tilde{M}}_k}_T|}{|\expval*{\hat{\tilde{M}}_k}_T|}\, .
\end{equation}
We first see that all moments have similar accuracy. Still, while not shown here, the moments do seem to converge individually when $M$ is increased. We wish to emphasize that the moments are very accurate with and without zero padding and thus must be used with caution as convergence criteria. 

\begin{figure}[t]
     \includegraphics[width=0.99\columnwidth]{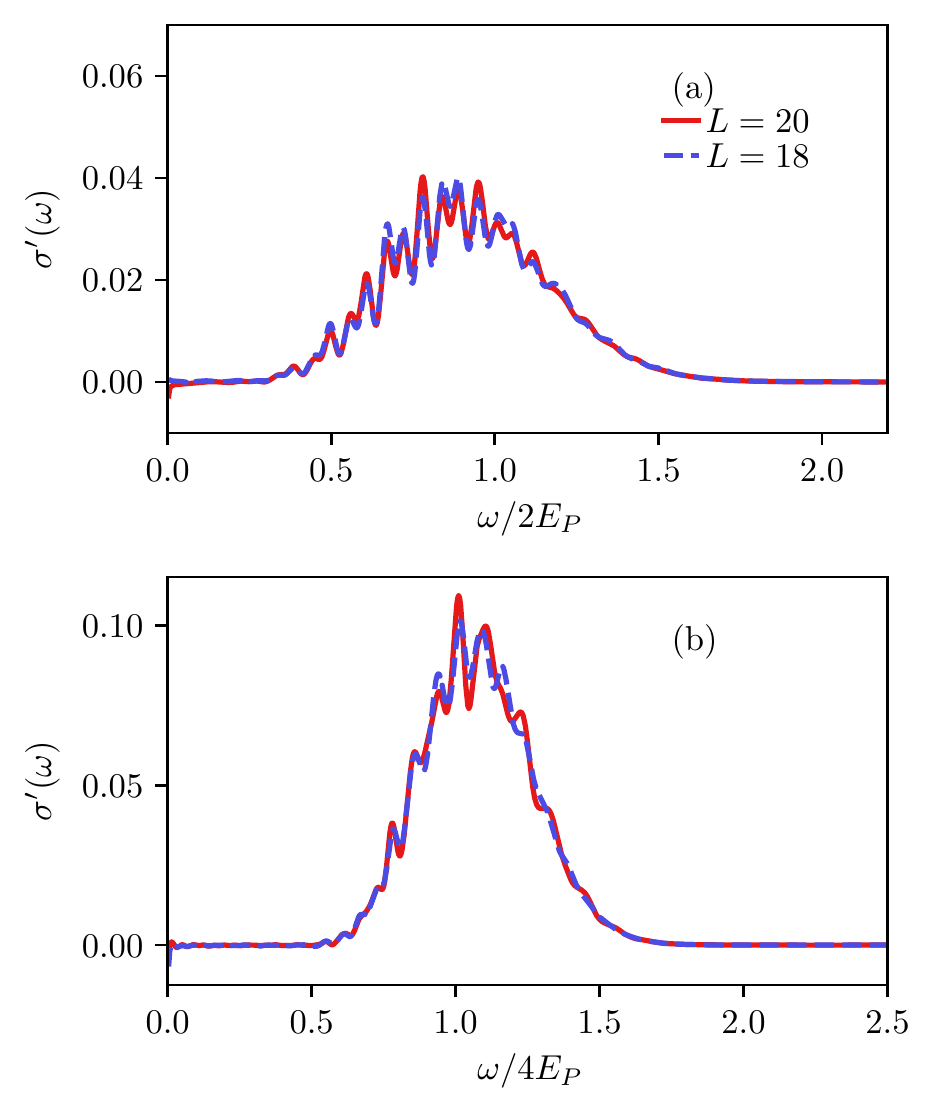}
\caption{(a) Ground-state optical conductivity for the polaron from Fig.~\ref{fig:OPCFT30}(b) for two different system sizes.  All other parameters are the same as in Fig.~\ref{fig:OPCFT30}(b). (b) Ground-state optical conductivity for the bipolaron from Fig.~\ref{fig:OPCFTbipol20}(c) for the same system sizes as in (a).  All other parameters are the same as in Fig.~\ref{fig:OPCFTbipol20}(c).}
\label{fig:dL}
\end{figure}

One sees the influence of the different $M$ on the optical conductivity in Fig.~\ref{fig:moms_comp}(c). Even for the largest $M$ used here, small differences in some of the peak amplitudes can be observed.  Still, the spectrum is quantitatively well captured. We also want to note that the real part of the optical conductivity in Fig.~\ref{fig:moms_comp}(c) has an $\eta$ dependence as well. This can be understood by inspecting the current-current correlation function (not shown here). There are still some $M$ dependent oscillations at longer times which get suppressed when $\eta$ is large. This is not the case for $t_{\textrm{ph}}/ \omega_0 \neq 0$, where the correlation functions have no notable oscillations for large times $t/\omega_0$. Similar behavior is seen for the ground-state calculations in Fig.~\ref{fig:OPCGS30}.  In some cases, we also observe some $L$ dependence in the amplitudes of the optical conductivities. This is illustrated in Fig.~\ref{fig:dL}.  

We further want to remark that the use of zero padding is not strictly necessary when using single-site TDVP as is done for $T/\omega_0=1$. This is because the computational cost does not increase with time. However, we do this to justifiably compare it to the ground-state data, for which we use p2TDVP-LBO, and which has an increased computational cost with time. We can conclude that for our data,  the convergence of the integral and the expectation values is a necessary but not a sufficient criterion to ensure proper convergence of the optical conductivity for parameters of the initial state. The convergence of the moments relative to each other seems to work better.   

Furthermore, the moments and the kinetic energy are used to ensure that the initial thermal state is converged with respect to $\rho_{\rm{LBO}}$, $\rho_{\rm{bond}}$, $M$, and the imaginary-time step. The largest relative difference is of order $\mathcal{O}(10^{-3})$ and is found for some quantities for the large coupling polaron and bipolaron at $T/\omega_0=1$. We also computed the relative overlap between states with different $\rho_{\rm{LBO}}$, $\rho_{\rm{bond}}$ and imaginary time steps and found a maximum difference from one of the order of $\mathcal{O}(10^{-5})$ for strong coupling and $T/\omega_0=1.0$ and $\mathcal{O}(10^{-6})$ for the rest of our finite $T/\omega_0$ data.

\section{Born-Oppenheimer surfaces} \label{sec:app3}
In order to derive an analytic expression for the real part of the optical conductivity for strong electron-phonon coupling, we first obtain the Born-Oppenheimer Hamiltonian~\cite{Born_54,Born_27} from Eq.~\eqref{eq:def_BOHam}.
For a fixed $\bar{q}$ this becomes a $2\cross 2 $ matrix,  in the polaron case,  which can be diagonalized \begin{equation} \label{eq:def_matrix1}
H_{\textrm{BO}}=\begin{pmatrix} \frac{\bar{q}^2}{2}\tilde{\omega}_0  + {\gamma}\bar{q} &-{t}_0  \\ -{t}_0 & \frac{\bar{q}^2}{2}\tilde{\omega}_0  - {\gamma}\bar{q} \end{pmatrix}\,  ,
\end{equation}
where $\tilde{\omega}_{0}=\omega_0-t_{\textrm{ph}}$. The Hamiltonian has the energies
$E^{\textrm{BO}}_{\pm}=\frac{1}{2}(\bar{q}^2\tilde{\omega}_0 \pm  2\sqrt{({\gamma}\bar{q})^2 +{t}_0^2} \, )$.  These energies give the Born-Oppenheimer surfaces. Their extrema with respect to $\bar{q}$ can be found by setting their derivatives to zero and one obtains $\bar{q}=0$, and $\bar{q}_{\textrm{min},\pm}=\pm \frac{\sqrt{\gamma^4-\tilde{\omega}_0^2{t}_0^2}}{\tilde{\omega}_0\gamma}$.  The Born-Oppenheimer surfaces can be seen in Fig.~\ref{fig:BOsurf_comb}(a) for different phonon dispersion relations. The minima  of the lower surface decrease as the phonon hopping goes from $t_{\rm{ph}}/\omega_0=-0.1$ to $+0.1$. 
We obtain $\Delta^{\textrm{BO}} =E^{\textrm{BO}}_{+}(q_{\textrm{min},\pm})-E^{\textrm{BO}}_{-}(q_{\textrm{min},\pm})=2 \sqrt{ \frac{\gamma^4 - \tilde{\omega}_0^2t_0^2}{\tilde{\omega}_0^2 }+t_0^2}$. For the bipolaron, we have a $4\cross 4$ matrix  
 \begin{equation} \label{eq:def_matrix2}
 H_{\textrm{BO}}=\begin{pmatrix} \frac{\bar{q}^2}{2}\tilde{\omega}_0+ 2{\gamma}\bar{q} &-t_0 &-t_0 &0 \\ -t_0 & \frac{\bar{q}^2}{2}\tilde{\omega}_0  & 0 & -t_0 \\-t_0 &  0 & \frac{\bar{q}^2}{2}\tilde{\omega}_0 & -t_0 \\
0 & -t_0 & -t_0 & \frac{\bar{q}^2}{2}\tilde{\omega}_0- 2{\gamma}\bar{q} 
\end{pmatrix}\,  ,
 \end{equation}
using the convention $\ket{\uparrow \downarrow,0},\ket{\uparrow,\downarrow},\ket{\downarrow,\uparrow},\ket{0,\uparrow \downarrow}$ for the electron occupation basis. Here, the surfaces are obtained by solving for the eigenvalues numerically. 

To obtain Eq.~\eqref{eq:def_analyt_formSC}, one uses the fact that the current operator connects the lowest and the first excited surfaces for a fixed $\bar{q}$.  To calculate the current-current correlation function, one must then compute~\cite{tokamkoff_22_2}
\begin{equation} \label{eq:def_formtoeval}
\begin{split}
\expval*{e^{i \hat H_g t}e^{-i \hat H_e t}}_T&=\expval*{e^{-\hat S}e^{\hat S}e^{i  \hat H_g t}e^{-\hat S}e^{\hat S}e^{-i \hat H_e t}e^{-\hat S}e^{\hat S}}_T\\&=\expval*{e^{i \hat H_e t}e^{\hat S}e^{-i \hat H_e t}e^{-\hat S}}_T=\expval*{e^{\hat S(t)}e^{-\hat S}}_T\,  ,
\end{split}
\end{equation}
where $\hat H_g$ is the harmonic oscillator Hamiltonian with frequencies $\tilde{\omega}_0=\omega_0-t_{\rm{ph}}$ and ladder operators $\hat{a}$ and $\hat{a}^{\dagger}$.  $\hat H_e$  is the harmonic oscillator Hamiltonian shifted with the distance $d$ and $e^{-\hat{S}}$ is the Lang-Firsov transformation operator, see Ref.~\cite{langfirsov}, with $\hat{S}=id\sqrt{\frac{1}{2}}(\hat a-\hat a^{\dagger})$. Evaluating Eq.~\eqref{eq:def_formtoeval}, see, e.g., Ref.~\cite{mahan_90},  rescaling with a factor of two for $L>2$, and expanding $e^{i\tilde{\omega}_0t}=1+i\tilde{\omega}_0t-\frac{1}{2}\tilde{\omega}_0^2t^2+\mathcal{O}(\tilde{\omega}_0^3)$ before the Fourier transformation leads to Eq.~\eqref{eq:def_analyt_formSC}.

 \bibliographystyle{biblev1}
 \bibliography{references}

\end{document}